\documentclass[12pt,epsf]{article}
\usepackage{amssymb,amsmath}
\usepackage{graphicx}
\usepackage[dvips]{color}

\newcommand{\be}{\begin{equation}}
\newcommand{\ee}{\end{equation}}
\newcommand{\bea}{\begin{eqnarray}}
\newcommand{\eea}{\end{eqnarray}}
\newcommand{\beas}{\begin{eqnarray*}}
\newcommand{\eeas}{\end{eqnarray*}}
\newcommand{\ba}{\begin{array}}
\newcommand{\ea}{\end{array}}

\newcommand{\nbox}{{\,\lower0.9pt\vbox{\hrule \hbox{\vrule height 0.2 cm \hskip 0.19 cm \vrule height 0.2 cm}\hrule}\,}}

\def\href#1#2{#2}
\input{tcilatex}

\begin{document}

\begin{titlepage}
\hfill
\vbox{
    \halign{#\hfil         \cr
           } 
      }  

\hbox to \hsize{{}\hss \vtop{ \hbox{}

}}

\begin{flushright}
 FPAUO-11-12
\end{flushright}

\vspace*{20mm}
\begin{center}

{\large \textbf{Young diagrams, Brauer algebras, and bubbling
geometries} }

{\Large \vspace{ 20mm} }

{\normalsize {Yusuke Kimura{\footnote{{\footnotesize e-mail:
kimurayusuke@uniovi.es}}} and   Hai Lin{\footnote{{\footnotesize
e-mail: hai.lin@usc.es}}}}  }

{\normalsize \vspace{10mm} }

{\normalsize \emph{${}^1$\textit{Departamento de Fisica,
Universidad de Oviedo,  \\
33007 Oviedo, Spain
}} }

{\normalsize \vspace{0.2cm} }

{\normalsize \emph{$^2$\textit{Department of Particle Physics,
Facultad de
Fisica, \\
Universidad de Santiago de Compostela, 15782 Santiago de
Compostela, Spain
\\
}} }

{\normalsize \vspace{0.2cm} }

%
%
%
%
%
%
%
%
\end{center}

\begin{abstract}

We study the $1/4$ BPS geometries corresponding to the $1/4$ BPS
operators of the dual gauge theory side, in ${\cal N}=4$ SYM. By
analyzing asymptotic structure and flux integration of the
geometries, we present a mapping between droplet configurations
arising from the geometries and Young diagrams of the Brauer
algebra. In particular, the integer $k$ classifying the operators
in the Brauer basis is mapped to the mixing between the two
angular directions.

\end{abstract}

\end{titlepage}

\vskip 1cm

\section{Introduction}

\vspace{1pt}\renewcommand{\theequation}{1.\arabic{equation}} %
\setcounter{equation}{0}

\vspace{1pt}

In this paper we study a particular problem in the context of the
gauge/gravity correspondence \cite{Maldacena:1997re},\cite{Gubser:1998bc},%
\cite{Witten:1998qj}. A family of 1/2 BPS geometries were found
and they are dual to a family of 1/2 BPS operators of
$\mathcal{N}$=$4$ SYM, as described in
\cite{Corley:2001zk},\cite{Berenstein:2004kk},\cite{Lin:2004nb}.
On the gauge side, they can be described by Shur polynomial
operators or Young diagrams. They can also be described by
wavefunctions of multi-body system. A droplet space on the gravity
side were found and the geometries dual to the corresponding
operators on the gauge side were mapped \cite{Lin:2004nb}. On the
gravity side, the geometries can be described by the phase space
of multi-body system and Young diagrams. These spacetime
geometries are nontrivial quantum states on the gravity side.

There are also 1/4 BPS operators and 1/8 BPS operators that are
also dual to corresponding 1/4 BPS geometries and 1/8 BPS
geometries. A general family of 1/4 BPS, 1/8 BPS geometries
corresponding to two-charge and three-charge
geometries were given in \cite{Donos:2006iy},\cite{Chen:2007du},\cite%
{Lunin:2008tf},\cite{Chong:2004ce}, pertaining to their
corresponding sectors. The conditions that the
$S^{3}$ shrinks or $S^{1}$ shrinks smoothly were analyzed in details in e.g. \cite%
{Chen:2007du},\cite{Lunin:2008tf}, \cite{Lin:2010nd}.

The careful analysis of the geometries of the gravity side shows
that the condition on the droplet space which characterizes the
regular geometries encodes the condition for having globally
well-defined spacetime geometries, as emphasized by
\cite{Lin:2004nb}. In this paper we also analyze these conditions
and characterize the geometries.

Meanwhile, gauge invariant operators which are dual to geometries have
scaling dimension of order $N^{2}$, which means that handling huge
combinatoric factors arising from summing up non-planar diagrams is
inevitable. A new observation is that the problem can be handled
systematically with the help of group theory. Following the earlier work
\cite{Corley:2001zk}, some bases for local gauge invariant operators with
two R-charges were given in \cite{Kimura:2007wy},\cite{Brown:2007xh},\cite%
{Bhattacharyya:2008rb}. These bases are labelled by Young diagrams, and the
calculations can be performed efficiently by representation theory. See \cite%
{Bhattacharyya:2008xy},\cite{Brown:2008ij},\cite{Kimura:2008ac},\cite%
{Koch:2008cm}, for example.

In this paper we study the relation between the droplet space of the
two-charge geometries or 1/4 BPS geometries and the dual two-charge
operators with the various bases. In particular we find the relation between
the droplet space and the basis built using elements of the Brauer algebra
given in \cite{Kimura:2007wy}. A class of the BPS operators were obtained at
weak coupling in \cite{Kimura:2010tx}, in which they are labelled by two
Young diagrams. We note that the size of each Young diagram is determined by
the R-charge of the fields and an integer. We will focus on how the Young
diagrams show up from the bubbling geometries. Other bases may also be
related to the droplet picture, since these bases can be related by
transformations from each other. There is also a droplet description from
other method on the gauge side by \cite{Berenstein:2005aa}.

The Young diagrams are also convenient for describing additional excitations
on these states. In particular, starting from a large dimension BPS operator
labelled by Young diagrams, one can modify the operators by replacing some
fields with other fields or multiplying some other fields. The presence of
those other fields in the operators make the states to be non-BPS and we can
describe those states as new excitations on the BPS states. See also related
discussions on those viewpoints, including e.g. \cite{Balasubramanian:2005mg}%
,\cite{Chen:2007gh},\cite{Koch:2008ah},\cite{Lin:2010sba},\cite%
{Carlson:2011hy}.

The organization of this paper is as follows. In section 2 and 3, we
introduce the general metric and flux. In section 4, we study the flux
integration on the droplet space. In section 5, we analyze metric functions
and mixings of metric components, for general droplet configurations. In
section 6, we study the large $R$ behavior of the mixings of the metric
components. In section 7, we analyze the configurations on the droplet space
and Young diagrams. In section 8, we discuss more about the operators
handled by the Brauer algebra. In section 9, we analyze the large $r$
asympotics of the geometry. Finally, in section 10, we briefly discuss our
results and conclusions. We also include several appendices.

\section{Metric and ansatz}

\renewcommand{\theequation}{2.\arabic{equation}} \setcounter{equation}{0}

\label{sec_metric_01}

\vspace{1pt}

We analyze two-charge geometries with $J_{1}$,$~J_{2}$ of two $U(1)$ global
symmetries inside $SO(6)$. General family of solutions have been studied in
\cite{Donos:2006iy},\cite{Chen:2007du},\cite{Lunin:2008tf},\cite%
{Chong:2004ce},\cite{Lin:2010nd}. They have $SO(4)\times SO(2)$
symmetry. The geometries have been studied from various
perspectives, see also e.g. \cite{Chong:2004ce}. It was found
\cite{Chen:2007du} that on the droplet space, the $S^{3}$ shrinks
smoothly, or the $S^{1}$ shrinks smoothly, see also e.g.
\cite{Lunin:2008tf},\cite{Lin:2010nd}.

They can be written via a K\"{a}hler potential
$K(z_{i},\bar{z}_{i};y)$. We have the 1/4 BPS geometry in the
form,
\begin{equation}
ds^{2}=-h^{-2}(dt+\omega )^{2}+h^{2}\left( dy^{2}+\frac{2\partial _{i}\bar{%
\partial}_{j}K}{Z+\frac{1}{2}}dz_{i}d\bar{z}_{j}\right) +ye^{G}d\Omega
_{3}^{2}+ye^{-G}d\psi ^{2},  \label{originalmetric}
\end{equation}%
and see Appendix \ref{appendix_gravity_a}.

\vspace{1pt} We will later analyze asymptotic structure of the geometries.
In order to perform the analysis, it is convenient to make a change of
coordinates
\begin{eqnarray}
y &=&r\cos \theta _{1},  \notag \\
z_{1} &=&R_{1}(r)\sin \theta _{1}\cos \theta _{2}e^{i\phi _{1}}, \\
z_{2} &=&R_{2}(r)\sin \theta _{1}\sin \theta _{2}e^{i\phi _{2}}.  \notag
\end{eqnarray}%
We also denote $\mu _{1}=\sin \theta _{1}\cos \theta _{2}$, $\mu _{2}=\sin
\theta _{1}\sin \theta _{2}$, $\mu _{3}=\cos \theta _{1}$ and $%
r_{i}=R_{i}(r)\mu _{i}$, $i=1,2$. We often use $R^{2}=r_{1}^{2}+r_{2}^{2}$.
We also define
\begin{equation}
S_{ij}=\frac{2e^{i(\phi _{i}-\phi _{j})}\partial _{i}\partial _{\bar{j}}K}{%
Z+1/2}.
\end{equation}

\vspace{1pt} In the following of this section, we mainly focus on the case
that $K=K(r_{1},r_{2},y)$. This means that the $S_{ij}$ is symmetric: $%
S_{ij}=S_{ji}$. Under this condition, with the shift of the angular
variables $\phi _{i}\rightarrow \phi _{i}-t$, the geometries can be
expressed by\vspace{1pt}\vspace{1pt}
\begin{eqnarray}
ds^{2} &=&-h^{-2}\left( 1+h_{ab}M_{a}M_{b}-S_{t}\right) dt^{2}+h^{2}\left(
\mu _{3}^{2}+S_{11}\mu _{1}^{2}T_{1}^{2}+S_{22}\mu
_{2}^{2}T_{2}^{2}+2S_{12}\mu _{1}\mu _{2}T_{1}T_{2}\right) dr^{2}  \notag \\
&+&2h^{2}(S_{11}R_{1}T_{1}\mu _{1}+S_{12}T_{2}\mu _{2}R_{1}-\mu _{1}r)d\mu
_{1}dr+2h^{2}(S_{22}R_{2}T_{2}\mu _{2}+S_{12}T_{1}\mu _{1}R_{2}-\mu
_{2}r)d\mu _{2}dr  \notag \\
&+&\sqrt{\Delta }r^{2}d\Omega _{3}^{2}+\frac{1}{\sqrt{\Delta }}\left( d\mu
_{3}^{2}+H_{1}d\mu _{1}^{2}+H_{2}d\mu _{2}^{2}\right) +2h^{2}\left(
S_{12}R_{1}R_{2}-\frac{\mu _{1}\mu _{2}}{\Delta }\right) d\mu _{1}d\mu _{2}
\notag \\
&+&\frac{\mu _{3}^{2}}{\sqrt{\Delta }}d\psi ^{2}+h^{-2}h_{ij}(d\phi
_{i}+M_{i}dt)(d\phi _{j}+M_{j}dt),  \label{geometries_new_variable_01}
\end{eqnarray}%
where we have defined $T_{i}=dR_{i}/dr$. We present the details of this
calculation in Appendix \ref{appendix_derivation_metric}.

The metric functions are defined as follows, \vspace{1pt}%
\begin{equation}
\Delta =\frac{\mu _{3}^{2}}{r^{2}}\frac{1+2Z}{1-2Z},
\end{equation}%
\begin{equation}
h^{-2}=\frac{r^{2}\Delta +\mu _{3}^{2}}{\sqrt{\Delta }},
\end{equation}%
\begin{equation}
H_{i}=\sqrt{\Delta }h^{2}\left( S_{ii}R_{i}^{2}-\frac{\mu _{i}^{2}}{\Delta }%
\right) .
\end{equation}%
The mixing between time and angles, which will play an important role to
determine the angular momenta of the geometries, is given by
\begin{eqnarray}
M_{1} &=&-1+\frac{-S_{2}\omega _{\phi _{1}}+N_{12}\omega _{\phi _{2}}}{%
S_{1}S_{2}-N_{12}^{2}},  \label{M_1_1} \\
M_{2} &=&-1+\frac{-S_{1}\omega _{\phi _{2}}+N_{12}\omega _{\phi _{1}}}{%
S_{1}S_{2}-N_{12}^{2}}.  \label{M_1_2}
\end{eqnarray}%
The functions in the angular part are
\begin{equation}
h_{11}=S_{1},\quad h_{22}=S_{2},\quad h_{12}=N_{12},
\end{equation}%
and the function in time is
\begin{equation}
S_{t}=S_{1}+S_{2}+2N_{12}+2\omega _{\phi _{1}}+2\omega _{\phi _{2}},
\end{equation}%
where
\begin{eqnarray}
&&S_{i}=h^{4}S_{ii}r_{i}^{2}-\omega _{\phi _{i}}^{2},~\quad i=1,2, \\
&&N_{12}=h^{4}S_{12}r_{1}r_{2}-\omega _{\phi _{1}}\omega _{\phi _{2}}.
\end{eqnarray}

The $AdS_{5}\times S^{5}$ can be recovered by plugging $\Delta =1$ and $%
R_{1}=R_{2}=\sqrt{r^{2}+r_{0}^{2}}$ in (\ref{geometries_new_variable_01}):
\begin{equation}
ds^{2}=-\left( 1+r^{2}\right)
dt^{2}+\frac{1}{1+r^{2}}dr^{2}+r^{2}d\Omega
_{3}^{2}+\sum_{i=1}^{3}\left( d\mu _{i}^{2}+\mu _{i}^2 d\phi
_{i}^{2}\right) ,
\end{equation}%
where the above are written in unit $r_{0}=1$, and we have renamed $\psi
=\phi _{3}$.

\section{Flux in general form}

\renewcommand{\theequation}{3.\arabic{equation}} \setcounter{equation}{0}

\vspace{1pt}

\label{sec_flux_01}\vspace{1pt}

\vspace{1pt}

\vspace{1pt}We start by writing out the metric, in particular by expanding
out the fibration over the time direction:
\begin{eqnarray}
ds_{10}^{2} &=&-h^{-2}\left( dt+\frac{1}{y}\left( \bar{\partial}_{i}\partial
_{y}K\right) d\bar{z}^{i}-\frac{1}{y}\left( \partial _{i}\partial
_{y}K\right) dz^{i}\right) ^{2}  \notag \\
&&+h^{2}\left( dy^{2}+\frac{2}{Z+\frac{1}{2}}\partial _{i}\bar{\partial}%
_{j}Kdz^{i}d\bar{z}^{j}\right) +y\left( e^{G}d\Omega _{3}^{2}+e^{-G}d\psi
^{2}\right) ,
\end{eqnarray}%
and $Z=\frac{1}{2}\mathrm{\tanh }G=-\frac{1}{2}y\partial _{y}(\frac{1}{y}%
\partial _{y}K)$.$~$The five form, is then given by
\begin{equation}
F_{5}=\left( -d\left( y^{2}e^{2G}(dt+\omega )\right) -y^{2}d\omega
+2i\partial _{i}\bar{\partial}_{j}Kdz^{i}d\bar{z}^{j}\right) \wedge d\Omega
_{3}+dual.
\end{equation}

There are different types of components. We can split them into two types of
components. The five form here may be written as
\begin{equation}
F_{5}=F_{2}\wedge d\Omega _{3}+F_{4}\wedge d\psi .
\end{equation}

The various components of $F_{2}$ are
\begin{eqnarray}
F_{2} &=&\partial _{y}\left( y^{2}e^{2G}\right) dt\wedge dy+\partial
_{i}\left( y^{2}e^{2G}\right) dt\wedge dz^{i}+\bar{\partial}_{i}\left(
y^{2}e^{2G}\right) dt\wedge d\bar{z}^{i}  \notag \\
&+&\left( iy(e^{2G}+1)\partial _{i}Z-\frac{i}{2y}\partial _{y}\left(
y^{2}e^{2G}\right) \partial _{i}\partial _{y}K\right) dz^{i}\wedge dy  \notag
\\
~ &-&\left( iy(e^{2G}+1)\bar{\partial}_{i}Z-\frac{i}{2y}\partial _{y}\left(
y^{2}e^{2G}\right) \bar{\partial}_{i}\partial _{y}K\right) d\bar{z}%
^{i}\wedge dy  \notag \\
&+&\frac{1}{2}iy\left( \partial _{i}\left( e^{2G}\right) \partial
_{j}\partial _{y}Kdz^{i}\wedge dz^{j}-\bar{\partial}_{i}\left( e^{2G}\right)
\bar{\partial}_{j}\partial _{y}Kd\bar{z}^{i}\wedge d\bar{z}^{j}\right)
\notag \\
&+&\left( 2i\partial _{i}\bar{\partial}_{j}K-iy\left( (e^{2G}+1)\left(
\partial _{i}\bar{\partial}_{j}\partial _{y}K\right) +\frac{1}{2}\left(
\partial _{i}\left( e^{2G}\right) \bar{\partial}_{j}\partial _{y}K+\bar{%
\partial}_{j}\left( e^{2G}\right) \partial _{i}\partial _{y}K\right) \right)
\right) dz^{i}\wedge d\bar{z}^{j}.  \notag \\
&&
\end{eqnarray}%
These components are multiplied by $d\Omega _{3}$. \newline

We write the full dual field strength for the five form as \cite{flexp}: {%
\fontsize{9}{11}\selectfont%
\begin{eqnarray}
F_{4} &=&\frac{1}{(1+2Z)^{2}}\left( y\partial _{j}\bar{\partial}%
_{k}K\partial _{y}\left( \frac{1}{y^{2}}\partial _{y}\partial _{i}K\partial
_{y}\bar{\partial}_{l}K\right) -\frac{1-4Z^{2}+2y\partial _{y}Z}{y^{2}}%
\partial _{j}\bar{\partial}_{l}K\partial _{i}\bar{\partial}_{k}K\right)
dz^{i}\wedge dz^{j}\wedge d\bar{z}^{k}\wedge d\bar{z}^{l}  \notag \\
&+&\frac{iy}{8(1+2Z)^{2}}\wedge \left( \partial _{i}\bar{\partial}%
_{j}K\partial _{y}\left( \frac{1}{y}\partial _{y}\bar{\partial}_{k}K\right)
dt\wedge dz^{i}\wedge d\bar{z}^{j}\wedge d\bar{z}^{k}-\partial _{j}\bar{%
\partial}_{i}K\partial _{y}\left( \frac{1}{y}\partial _{y}\partial
_{k}K\right) dt\wedge dz^{j}\wedge d\bar{z}^{k}\wedge d\bar{z}^{i}\right)
\notag \\
&+&\frac{(1-2Z)}{2y^{2}(1+2Z)}\left( \partial _{i}\bar{\partial}_{k}K\left(
\frac{1}{8}\left( \frac{1-2Z}{2y}\partial _{j}\left( y^{2}e^{2G}\right)
-\partial _{j}\partial _{y}K\right) +\frac{2y\epsilon ^{ef}\epsilon
^{gl}\partial _{f}\bar{\partial}_{l}K\partial _{y}\partial _{e}K\partial
_{y}\partial _{j}\bar{\partial}_{g}K}{(1-2Z)det\left( \partial _{i}\bar{%
\partial}_{j}K\right) }\right) dz^{i}\wedge dz^{j}\wedge d\bar{z}^{k}\wedge
dy\right.  \notag \\
&-&\left. \partial _{i}\bar{\partial}_{k}K\left( \frac{1}{8}\left( \frac{1-2Z%
}{2y}\bar{\partial}_{j}\left( y^{2}e^{2G}\right) -\bar{\partial}_{j}\partial
_{y}K\right) +\frac{2y\epsilon ^{ef}\epsilon ^{gl}\partial _{l}\bar{\partial}%
_{f}K\partial _{y}\bar{\partial}_{e}K\partial _{y}\partial _{g}\bar{\partial}%
_{j}K}{(1-2Z)det\left( \partial _{i}\bar{\partial}_{j}K\right) }\right)
dz^{i}\wedge d\bar{z}^{j}\wedge d\bar{z}^{k}\wedge dy\right)  \notag \\
&+&\frac{i(1-2Z)}{y(1+2Z)}\left( \frac{\partial _{i}\bar{\partial}_{j}K}{8}+%
\frac{2y}{1-2Z}\left( \frac{\partial _{y}\partial _{i}\bar{\partial}_{j}K}{16%
}+\frac{\epsilon ^{ek}\epsilon ^{lf}\partial _{k}\bar{\partial}_{l}K\partial
_{i}\bar{\partial}_{j}K\partial _{y}\partial _{e}\bar{\partial}_{f}K}{%
det\left( \partial _{i}\bar{\partial}_{j}K\right) }\right) \right) dt\wedge
dz^{i}\wedge d\bar{z}^{j}\wedge dy  \notag \\
&&  \label{flux_02}
\end{eqnarray}%
}where these should be multiplied by $d\psi $.

\vspace{1pt}

\section{Flux integration}

\vspace{1pt}\renewcommand{\theequation}{4.\arabic{equation}} %
\setcounter{equation}{0}

\label{sec_flux_integration_01}

We focus on the droplet space on which the $S^{3}$ or $S^{1}~$vanishes.
These occur at $y=0$. The droplet space is divided into two droplet regions,
with one region where $Z=-\frac{1}{2}$ and the $S^{3}$ vanishes, and another
region where $Z=\frac{1}{2}$ and the $S^{1}$ vanishes. The flux integration
at $y=0$ involves several different situations, depending on different types
of droplets.

We look at the first term in the expression (\ref{flux_02}). We can find the
$Z=-\frac{1}{2}$ small $y$ behavior of this term and find this term near $%
y=0 $, with $Z=-\frac{1}{2}$ droplet region,
\begin{equation}
-dz^{1}\wedge d\bar{z}^{1}\wedge dz^{2}\wedge d\bar{z}^{2}\wedge d\psi .
\end{equation}%
The flux integration in $Z=-\frac{1}{2}$ is%
\begin{equation}
\int_{M_{4}\times S^{1}}F_{5}=\int_{M_{4}\times S^{1}}dz^{1}\wedge
dz^{2}\wedge d\bar{z}^{1}\wedge d\bar{z}^{2}\wedge d\psi =2\pi
\int_{M_{4}}dz^{1}\wedge dz^{2}\wedge d\bar{z}^{1}\wedge d\bar{z}^{2},
\end{equation}%
with
\begin{equation}
\frac{1}{8\pi ^{3}l_{p}^{4}}\int_{M_{4}}dz^{1}\wedge dz^{2}\wedge d\bar{z}%
^{1}\wedge d\bar{z}^{2}=N_{i},  \label{flux_N_i}
\end{equation}%
which are quantized, due to the quantization of the $F_{5}$ flux. $N_{i}$ is
the flux quantum number in each region. The volume of $\psi ~$is $%
V_{S^{1}}=2\pi .$

\vspace{1pt} For example, for $AdS_{5}\times S^{5},~$when
$Z=-\frac{1}{2},$
we have $r=0$, so $z_{1}=r_{0}\sin \theta _{1}\cos \theta _{2}e^{i\phi _{1}}$%
, \newline
$z_{2}$$=r_{0}\sin \theta _{1}\sin \theta _{2}e^{i\phi _{2}}$, and
\begin{eqnarray}
\frac{1}{4}\int_{M_{4}}dz^{1}\wedge dz^{2}\wedge d\bar{z}^{1}\wedge d\bar{z}%
^{2} &=&\frac{\pi ^{2}}{2}r_{0}^{4}, \\
\frac{1}{8\pi ^{3}l_{p}^{4}}\int_{M_{4}}dz^{1}\wedge dz^{2}\wedge d\bar{z}%
^{1}\wedge d\bar{z}^{2} &=&\frac{r_{0}^{4}}{4\pi l_{p}^{4}}=N,
\label{flux_a_02}
\end{eqnarray}%
where for the ground state, $M_{4}$ is the compact region bounded by $%
\left\vert z_{1}\right\vert ^{2}+\left\vert z_{2}\right\vert ^{2}=r_{0}^{2}$%
.~We have $dz^{1}\wedge d\bar{z}^{1}\wedge dz^{2}\wedge d\bar{z}%
^{2}=-4dx_{1}\wedge dx_{2}\wedge dx_{3}\wedge dx_{4},$ where $%
z_{1}=x_{1}+ix_{2},z_{2}=x_{3}+ix_{4}$. %
We may view this as a 4d droplet space.

The flux quantum number would map to the lengths $N_{i}$ of the vertical
edges of the Young diagram operators. Here $i$ denotes the different $Z=-%
\frac{1}{2}$ regions at $y=0$. Compact $Z=-\frac{1}{2}~$droplets correspond
to $AdS_{5}$ asymptotics, while non-compact $Z=-\frac{1}{2}~$droplets could
give rise to other asymptotics.

Now we look at the flux integration in $Z=\frac{1}{2}$ region. On the other
hand, at $y=0$, with $Z=\frac{1}{2}$ droplet region, the $F_{5}$ has
components
\begin{equation}
2i\partial _{i}\bar{\partial}_{j}K_{0}dz^{i}\wedge d\bar{z}^{j}\wedge
d\Omega _{3},
\end{equation}%
where $K_{0}$ is defined in (\ref{K_01_02}) in the appendix, so we have the
flux integrations
\begin{align}
\int_{D_{2}\times S^{3}}F_{5}& =\int_{D_{2}}4\pi ^{2}i\partial _{i}\bar{%
\partial}_{j}K_{0}dz^{i}\wedge d\bar{z}^{j}, \\
\int_{\tilde{D}_{2}\times S^{3}}F_{5}& =\int_{\tilde{D}_{2}}4\pi
^{2}i\partial _{i}\bar{\partial}_{j}K_{0}dz^{i}\wedge d\bar{z}^{j},
\end{align}%
where $V_{S^{3}}=2\pi ^{2}$, and $D_{2},~\tilde{D}_{2}$ are two dimensional
domains. We have that
\newline
\begin{eqnarray}
\frac{1}{(2\pi )^{2}l_{p}^{4}}\int_{D_{2}}i\partial _{i}\bar{\partial}%
_{j}K_{0}dz^{i}\wedge d\bar{z}^{j} &=&m_{i}, \\
\frac{1}{(2\pi )^{2}l_{p}^{4}}\int_{\tilde{D}_{2}}i\partial _{i}\bar{\partial%
}_{j}K_{0}dz^{i}\wedge d\bar{z}^{j} &=&{\tilde{m}}_{i}.
\end{eqnarray}%
Here, we have non-contractible two-cycles, so we have several situations.

For $AdS_{5}\times S^{5}$,
\newline
\begin{equation}
K_{0}=\frac{1}{2}aR^{2}-\frac{1}{2}q\log (aR^{2}).  \label{ads_expression}
\end{equation}%
In that case there is no such two-cycles. We have $\partial _{1}\bar{\partial%
}_{1}K_{0}|_{z_{2}=0}=\frac{1}{2}a,\partial _{2}\bar{\partial}%
_{2}K_{0}|_{z_{1}=0}=\frac{1}{2}a,$ so the flux components along the $z_{1}~$%
plane where $z_{2}=0$ is constant, and the flux components along
the $z_{2}$ plane where $z_{1}=0$ is also constant.

\begin{figure}[h]
\begin{center}
\includegraphics[width=11.5cm]{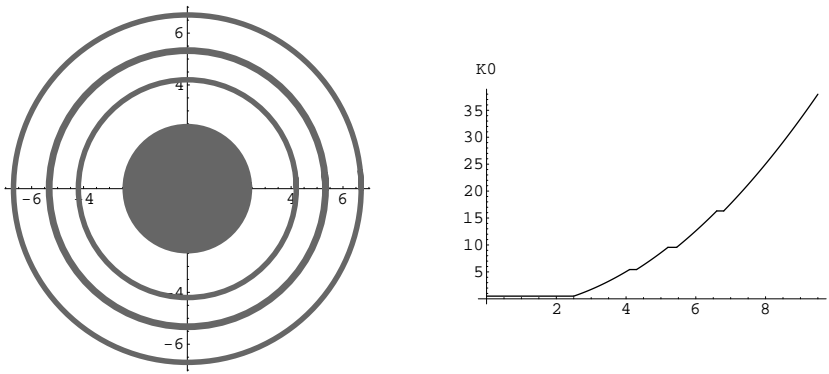}
\end{center}
\caption{{\protect\small On the left side is the droplet configuration on
the $z_1$ plane. On the right side is the plot of $K_0$ on the $z_1$ plane
along the radial axis. Similarly for the $z_2$ plane. There are thin $Z=-%
\frac{1}{2}$ regions. In the limit that the thin black strips go to zero,
the function goes to that of the AdS expression. }}
\label{plot_01_01}
\end{figure}

One of the simplest situations is that there are many $Z=-\frac{1}{2}~$%
regions on the $z_{1}~$plane and $z_{2}~$plane. We can form the two-cycles
between the $Z=-\frac{1}{2}~$regions on the $z_{1}~$plane for $D_{2}$, and
on the $z_{2}~$plane for $\tilde{D}_{2}$. They are the white strips in
figure \ref{plot_01_01}. Now we consider very thin $Z=-\frac{1}{2}~$regions
on top of the $Z=\frac{1}{2}$ background. See figure \ref{plot_01_01} for an
example of thin $Z=-\frac{1}{2}~$regions. In this case, on the $Z=\frac{1}{2}
$ background, away from the very thin $Z=-\frac{1}{2}~$regions, $\partial
_{1}\bar{\partial}_{1}K_{0}|_{z_{2}=0}$, $\partial _{2}\bar{\partial}%
_{2}K_{0}|_{z_{1}=0}$ are approximately given by those of the AdS expression
e.g. (\ref{ads_expression}).

We now consider the $D_{2}$ domain of $Z=\frac{1}{2}$ on $z_{1}$ plane at $%
z_{2}=0$, surrounded by $Z=-\frac{1}{2}$ regions, and we have
\vspace{3pt}
\begin{equation}
\frac{1}{(2\pi )^{2}l_{p}^{4}}\int_{D_{2}}i\partial _{1}\bar{\partial}%
_{1}K_{0}dz^{1}\wedge d\bar{z}^{1}=m_{i}.
\end{equation}%
\vspace{2.5pt}
Similarly $\tilde{D}_{2}$ is the domain of $Z=\frac{1}{2}$ on $z_{2}~$%
plane~at $z_{1}=0,$ surrounded by $Z=-\frac{1}{2}~$regions, and we
have
\vspace{3pt}
\begin{equation}
\frac{1}{(2\pi )^{2}l_{p}^{4}}\int_{\tilde{D}_{2}}i\partial _{2}\bar{\partial%
}_{2}K_{0}dz^{2}\wedge d\bar{z}^{2}={\tilde{m}}_{i}.
\end{equation}%
\vspace{2pt}

According to the BPS operators in \cite{Kimura:2010tx} expressed
by the Brauer basis in \cite{Kimura:2007wy}, the operators can be
labelled by two Young diagrams. These operators are briefly summarized in section \ref%
{sec_operater_side}. We may identify the flux quantum numbers $m_{i}$ with
the size of the horizontal edges $m_{i}$ of the first Young diagram, and
identify the flux quantum number$\ {\tilde{m}}_{i}$ with the size of the
horizontal edges ${\tilde{m}}_{i}~$of the second Young diagram. The two
Young diagrams have total number of boxes $m-k$ and $n-k$ respectively,
\newline

\begin{eqnarray}
\sum_{i}\sum_{i^{\prime }=1}^{i}m_{i^{\prime }}N_{i} &=&m-k,
\label{totalboxinm} \\
\sum_{i}\sum_{i^{\prime }=1}^{i}{\tilde{m}}_{i^{\prime }}\tilde{N}_{i}
&=&n-k.  \label{totalboxinn}
\end{eqnarray}%
The flux integrations at $y=0$ show how the edges of Young diagrams would be
mapped, according to above discussions.

\vspace{1pt}Following the earlier works, we further identify the appropriate
variables and functions to characterize the geometries, and focused on the
droplet space which are divided into two droplet regions, with $Z=-\frac{1}{2%
}$ where the $S^{3}$ vanishes and with $Z=\frac{1}{2}$ where the $S^{1}$
vanishes. The flux quantum numbers are determined by the flux integrals. The
dimension of the operator is $m+n$. Note that these are the flux
quantization in the small $y$ region. These are not the same as the flux
quantization in the large $r$ region.

\vspace{1pt}$K_{0}$ plays an important role in these flux integrations that
involve the two-cycles in the $Z=\frac{1}{2}$ regions. $K_{0}$ can be solved
by the coupled equations (\ref{coupledeqK0K1}) of $K_{0},K_{1}$ for these
regions. Figure \ref{plot_01_01} shows a plot of $K_{0}$ where there are
several thin $Z=-\frac{1}{2}$ strips. 

\vspace{1pt}
\vspace{1pt}

\section{Droplet space, metric functions and $N_{1 2}$, $M_ 1$, $M_ 2$}

\renewcommand{\theequation}{5.\arabic{equation}} \setcounter{equation}{0}

\label{sec_K_expansion_N12_M1_M2_general}

\subsection{Metric functions}

\label{smallyexpansion_metricfunction}

\vspace{1pt}We now study the metric functions in the geometry as well as the
metric components mixing time and angles, as well as components mixing the
angles. The expressions in this section are valid for all possible ranges of
$z_{i}$, $\bar{z}_{i}$. In a different section \ref{sec_N12_M1_M2_R_large},
we will study their large $R$ behaviors, in the large $R$ region. We denote $%
R^{2}=r_{1}^{2}+r_{2}^{2}$\vspace{1pt}, and $r_{1}=\left\vert
z_{1}\right\vert ,r_{2}=\left\vert z_{2}\right\vert .$

We expand $K$ in powers of $y^{2}$.~The entire solutions to $K$ are
determined by the functions $K_{0},K_{1},K_{2},$ see appendix \ref%
{appendix_gravity_a}. The equations for $K$ are in Appendix \ref%
{appendix_gravity_a}.

In this section we study exact expressions in all possible ranges of $z_{i}$%
, $\bar{z}_{i}$, but for small $y$, since we consider $K_{0},K_{1},K_{2}$ in
the series expansion in powers of $y^{2}$.

Near $Z=1/2$, we have the expansion
\begin{eqnarray}
K &=&-\frac{1}{4}y^{2}\log
y^{2}+K_{0}+y^{2}K_{1}+y^{4}K_{2}+y^{6}K_{3}+O(y^{8}), \\
Z &=&\frac{1}{2}-4y^{2}K_{2}-12y^{4}K_{3}+O(y^{6}).  \label{expandZat1/2}
\end{eqnarray}%
These expressions will be used in the later derivations in this section.

We also use
\begin{equation}
\omega _{\phi _{i}}=-\frac{1}{2y}\partial _{y}(r_{i}\partial
_{r_{i}}K)=-r_{i}\partial _{r_{i}}K_{1}-2y^{2}r_{i}\partial
_{r_{i}}K_{2}+O(y^{4}).
\end{equation}

The $\Delta $ can be expanded as\vspace{1pt}%
\begin{eqnarray}
\Delta &=&\frac{\mu _{3}^{2}}{r^{2}}\frac{1+2Z}{1-2Z}  \notag
\label{deltaexpansion} \\
&=&\frac{\mu _{3}^{2}}{r^{2}}\frac{1-4y^{2}K_{2}-12y^{4}K_{3}+O(y^{6})}{%
4y^{2}K_{2}+12y^{4}K_{3}+O(y^{6})}  \notag \\
&=&\frac{1}{4r^{4}K_{2}}\left( 1-4y^{2}K_{2}-3y^{2}\frac{K_{3}}{K_{2}}%
+O(y^{4})\right) .
\end{eqnarray}%
Note that $y=r\mu _{3}$.

\vspace{1pt}Around $y=0$ with $Z=\frac{1}{2}$, which is approached by $\mu
_{3}=0$, we may expand it as
\begin{equation}
Z=\frac{1}{2}\frac{1-\frac{r_{0}^{2}\mu _{3}^{2}}{r^{2}\Delta }}{1+\frac{%
r_{0}^{2}\mu _{3}^{2}}{r^{2}\Delta }}=\frac{1}{2}-\frac{r_{0}^{2}y^{2}}{%
r^{4}\Delta }+\left( \frac{r_{0}^{2}y^{2}}{r^{4}\Delta }\right)
^{2}+O(y^{6}).  \label{Z_exp_delta}
\end{equation}%
Comparing (\ref{expandZat1/2}),(\ref{Z_exp_delta}), we have
\begin{equation}
K_{2}=\frac{r_{0}^{2}}{4r^{4}\Delta }+O(y^{2}),
\end{equation}%
\begin{equation}
h^{2}=\sqrt{4K_{2}}+O(y^{2}).
\end{equation}

\subsection{Mixing components}

We study particularly the important functions $M_{1},M_{2},N_{12}$ which are
the mixing between time and the angles, and the mixing between the angles
themselves.

We focus on the components of the geometry,%
\begin{equation}
\frac{h_{ij}}{h^{2}}(d\phi _{i}+M_{i}dt)(d\phi _{j}+M_{j}dt).
\end{equation}%
In particular, there is also a mixing term between the angles,
\begin{equation}
2\frac{N_{12}}{h^{2}}(d\phi _{1}+M_{1}dt)(d\phi _{2}+M_{2}dt),
\end{equation}%
where $h_{12}=N_{12}$.

$M_{1},M_{2}~$account for the mixing between $t$ and $\phi _{1},\phi _{2}$
respectively. The function $N_{12}~$account for the mixing of angles $\phi
_{1}$ and $\phi _{2}.$

Using the small $y$ expansions presented in section \ref%
{smallyexpansion_metricfunction} and $h^{2}=(1/4-Z^{2})/y^{2}$, one can
expand the $S_{i}$, $N_{12}$ in section \ref{sec_metric_01} as
\begin{eqnarray}
S_{i} &=&h^{4}S_{ii}r_{i}^{2}-\omega _{\phi _{i}}^{2}  \notag \\
&=&2K_{2}\partial _{t_{i}}^{2}K_{0}-(\partial _{t_{i}}K_{1})^{2}+O(y^{2})
\notag \\
&=&s_{i}+O(y^{2})
\end{eqnarray}%
and
\begin{eqnarray}
N_{12} &=&h^{4}S_{12}r_{1}r_{2}-\omega _{\phi _{1}}\omega _{\phi _{2}}
\notag \\
&=&2K_{2}\partial _{t_{1}}\partial _{t_{2}}K_{0}-(\partial
_{t_{1}}K_{1})(\partial _{t_{2}}K_{1})+O(y^{2})  \notag  \label{N_12} \\
&=&n_{12}+O(y^{2}),
\end{eqnarray}%
where
\begin{eqnarray}
s_{1} &=&2K_{2}\partial _{t_{1}}^{2}K_{0}-(\partial _{t_{1}}K_{1})^{2}
\notag \\
s_{2} &=&2K_{2}\partial _{t_{2}}^{2}K_{0}-(\partial _{t_{2}}K_{1})^{2}
\notag \\
n_{12} &=&2K_{2}\partial _{t_{1}}\partial _{t_{2}}K_{0}-(\partial
_{t_{1}}K_{1})(\partial _{t_{2}}K_{1})  \label{def-of-sn}
\end{eqnarray}%
have been defined. We also have introduced $t_{i}=\log r_{i}$.\footnote{%
It would be convenient to use
\begin{equation*}
e^{i(\phi _{i}-\phi _{j})}\partial _{i}\partial _{\bar{j}}=\frac{1}{%
4r_{i}r_{j}}\partial _{t_{i}}\partial _{t_{j}}.
\end{equation*}%
} It is interesting to note that only the 2nd derivatives of $K_{0}$, and
1st derivatives of $K_{1}$, and no derivative of $K_{2}$ appears in the
expression.

Hence the mixing functions $M_{1}$ and $M_{2}$ in (\ref{M_1_1}),(\ref{M_1_2}%
) can be expanded as
\begin{eqnarray}
M_{1} &=&-1+\frac{s_{2}\partial _{t_{1}}K_{1}-n_{12}\partial _{t_{2}}K_{1}}{%
s_{1}s_{2}-n_{12}^{2}}+O(y^{2}),  \label{M_1} \\
M_{2} &=&-1+\frac{s_{1}\partial _{t_{2}}K_{1}-n_{12}\partial _{t_{1}}K_{1}}{%
s_{1}s_{2}-n_{12}^{2}}+O(y^{2}).  \label{M_2}
\end{eqnarray}%
These are exact expressions for any value of $r_{1},r_{2}.$ The special case
of large $R^{2}(=r_{1}^{2}+r_{2}^{2})$ will be discussed in section \ref%
{sec_N12_M1_M2_R_large}.

The mixing function between the angles will be particularly important in our
analysis to make a connection with the gauge theory. In general cases, it is
nonzero, $n_{12}\neq 0$. But there are special cases when $n_{12}=0$. In
this case, the $M_{i}$ becomes simpler. The special case that $n_{12}=0$
correspond to defining $N_{i}$:
\begin{equation}
N_{i}=M_{i}|_{n_{12}=0}.
\end{equation}%
We find that%
\begin{eqnarray}
N_{1} &=&-1+\frac{\partial _{t_{1}}K_{1}}{s_{1}}+O(y^{2}), \\
N_{2} &=&-1+\frac{\partial _{t_{2}}K_{1}}{s_{2}}+O(y^{2}).
\end{eqnarray}

\section{Droplet space and analysis of mixing components}

\vspace{1pt}\renewcommand{\theequation}{6.\arabic{equation}} %
\setcounter{equation}{0}

\vspace{1pt}\label{sec_N12_M1_M2_R_large}

\vspace{1pt} We have presented general expressions for small $y$
expansions, in section \ref{sec_K_expansion_N12_M1_M2_general}.
The geometries are parameterized by $y$ and $z_{i},\bar{z}_{i}$.
We denote $R^{2}=r_{1}^{2}+r_{2}^{2}$, where $r_{1}=\left\vert
z_{1}\right\vert ,r_{2}=\left\vert z_{2}\right\vert$. Now we
analyze these expressions in the large $R$ region, by expansions
in $1/R^{2}$. We find solutions in series expansion in powers of
$1/R^{2}$. We will analyze the behavior of the mixing components
$N_{12},M_{1},M_{2}$ in large $R$.

The K\"{a}hler potential is given by the Monge-Ampere equation (\ref%
{MA_general}), and in the small $y$ expansion, the equation gives a set of
equations for $K_{0}$ and $K_{1}$ \cite{Lin:2010nd},
\begin{equation}
(\partial _{t_{1}}\partial _{t_{1}}K_{0})(\partial _{t_{2}}\partial
_{t_{2}}K_{1})+(\partial _{t_{1}}\partial _{t_{1}}K_{1})(\partial
_{t_{2}}\partial _{t_{2}}K_{0})-2(\partial _{t_{1}}\partial
_{t_{2}}K_{0})(\partial _{t_{1}}\partial _{t_{2}}K_{1})=0,
\end{equation}%
\begin{equation}
(\partial _{t_{1}}\partial _{t_{1}}K_{0})(\partial _{t_{2}}\partial
_{t_{2}}K_{0})-(\partial _{t_{1}}\partial _{t_{2}}K_{0})^{2}=\frac{4}{e}%
e^{2t_{1}+2t_{2}}e^{2K_{1}},
\end{equation}%
where $t_{i}=\log r_{i}$.

This set of the equations have some rescaling transformations:
\begin{eqnarray}
K_{0} &\rightarrow &qK_{0}, \\
t_{i} &\rightarrow &\xi _{i}t_{i}+b_{i}~~\quad (r_{i}\rightarrow
e^{b_{i}}(r_{i})^{\xi _{i}}), \\
K_{1} &\rightarrow &K_{1}+\log q-\sum_{i}((\xi _{i}-1)t_{i}+b_{i}+\frac{1}{2}%
\log \xi _{i}^{2}).
\end{eqnarray}%
Overall constant shifts in $K_{0}$ are not important because they appear
with derivatives in the metric. The $\xi _{i}=-1$ transformation is the
inversion transformation in \cite{Lin:2010nd}, which exchanges small $r_{i}$
with large $r_{i}$.

The most general form of solutions representing the $AdS_{5}\times S^{5}$
geometry, in a region with $Z=1/2$, is \footnote{%
If we consider the following one,
\begin{equation*}
K_{0}=\frac{1}{2}(a(r_{1}^{2}+r_{2}^{2}))-\frac{q}{2}\log
(a(sr_{1}^{2}+r_{2}^{2})),
\end{equation*}%
the equations require $s=1$.}
\begin{equation}
K_{0}=\frac{1}{2}(a(sr_{1}^{m}+r_{2}^{n}))-\frac{q}{2}\log
(a(sr_{1}^{m}+r_{2}^{n})).
\end{equation}%
Using the rescaling transforms, this may be brought to the simplest form $%
K_{0}=\frac{1}{2}R^{2}-\frac{1}{2}\log R^{2}$, where $%
R^{2}=r_{1}^{2}+r_{2}^{2}$. In our analysis, parameters $a,q$ will be
reserved to account for the possible scaling transformation for $R$ (or $%
z_{i},\bar{z}_{i}$) and $K$.

\vspace{1pt}For example, we can have the rescaling transformations:%
\begin{eqnarray}
K_{0} &\rightarrow &qK_{0}, \\
\frac{q}{a}R^{2} &\rightarrow &R^{2},  \label{rescaling_R2} \\
sr_{1}^{2} &\rightarrow &r_{1}^{2},
\end{eqnarray}%
together with constant shifts of $K_{1}$, and
rescaling of $y$.

We are interested in more general solutions. We will first
consider from large $R$ point of view. In the large $R$, we have
that, $r^{2}=aR^{2}(1+O(1/r^{2}))+O(y^{2})$, so in the region
where $R$ is large, we have that $r$ is also large.

We denote $R^{2}=r_{1}^{2}+r_{2}^{2}$. Note that
$aR^{2},ar_{1}^{2},ar_{2}^{2}$ often appear in combinations in
their products, due to that $\sqrt{a}$ rescales the $r_{i}$
coordinates, as in (\ref{rescaling_R2}).

We find a family of expressions that satisfy the set of the equations, as
follows:

$K_{0}$ is given by
\begin{equation}
K_{0}=\frac{1}{2}aR^{2}-\frac{1}{2}q\log (aR^{2})+K_{0}^{(1)},  \label{K0}
\end{equation}%
\begin{eqnarray}
\partial _{t_{1}}^{2}K_{0} &=&2ar_{1}^{2}\left( 1-\frac{qr_{2}^{2}}{aR^{4}}%
+O(1/r^{4})\right) , \\
\partial _{t_{2}}^{2}K_{0} &=&2ar_{2}^{2}\left( 1-\frac{qr_{1}^{2}}{aR^{4}}%
+O(1/r^{4})\right) , \\
\partial _{t_{1}}\partial _{t_{2}}K_{0} &=&\frac{2qr_{1}^{2}r_{2}^{2}}{R^{4}}%
\left( 1+\frac{q\alpha _{0}}{aR^{2}}+O(1/r^{4})\right) ,  \label{K_alpha_0}
\end{eqnarray}
and $K_{1}$ is
\begin{equation}
K_{1}=\frac{1}{2}\log \left( \frac{a\left( aR^{2}-q\right) }{R^{2}}\right) +%
\frac{1}{2}+K_{1}^{(1)},
\end{equation}%
\begin{eqnarray}
\partial _{t_{1}}K_{1} &=&\frac{qr_{1}^{2}}{aR^{4}}\left( 1+\frac{q}{aR^{2}}%
(1+\kappa _{1})+O(1/r^{4})\right) ,  \label{K_kappa_1} \\
\partial _{t_{2}}K_{1} &=&\frac{qr_{2}^{2}}{aR^{4}}\left( 1+\frac{q}{aR^{2}}%
(1+\kappa _{2})+O(1/r^{4})\right) ,  \label{K_kappa_2}
\end{eqnarray}
and $K_{2}$ is
\begin{eqnarray}
K_{2} &=&\frac{q}{4(aR^{2}-q)^{2}}+K_{2}^{(1)} \\
&=&\frac{q}{4a^{2}R^{4}}\left( 1+\frac{q}{aR^{2}}(2+\alpha
_{2})+O(1/r^{4})\right) .  \label{K_alpha_2}
\end{eqnarray}

The expressions without $K_{0}^{(1)},K_{1}^{(1)},K_{2}^{(1)}$ is the
solution for AdS (see \cite{Lin:2010nd} for other related analysis). $%
K_{0}^{(1)},K_{1}^{(1)},K_{2}^{(1)}$ are deviations from AdS. The $\alpha
_{0}$, $\kappa _{1}$, $\kappa _{2}$, $\alpha _{2}~$are the effect of turning
on $K_{0}^{(1)},K_{1}^{(1)},K_{2}^{(1)}$. In the second lines in (\ref%
{K_alpha_0}),(\ref{K_kappa_1}),(\ref{K_kappa_2}),(\ref{K_alpha_2}),
the large $r$ expansion with the effect of
$K_{0}^{(1)},K_{1}^{(1)},K_{2}^{(1)}$ are given.

From the differential equations, we find a family of $K^{(1)}$,%
\begin{eqnarray}
K_{0}^{(1)} &=&\frac{q^{2}}{a}\left( \frac{dr_{1}^{2}+er_{2}^{2}}{R^{4}}%
+O(1/r^{4})\right) ,\quad  \label{K_0_(1)_d_e} \\
K_{1}^{(1)} &=&-\frac{q^{2}}{a^{2}R^{2}}\left( \frac{%
(d-e)(r_{1}^{2}-r_{2}^{2})}{R^{4}}+O(1/r^{4})\right) , \\
K_{2}^{(1)} &=&\frac{q^{2}}{4a^{3}R^{6}}\left( c-2(d+e)+\frac{%
6(d-e)(r_{1}^{2}-r_{2}^{2})}{R^{2}}\right) +\frac{q^{2}}{4a^{2}R^{4}}%
O(1/r^{4}).  \label{K_(1)_d_e}
\end{eqnarray}%
We note that three parameters $c$,$d$,$e$ have come in, and they will be
identified with the three parameters $m,n,k$ of Young diagrams in (\ref%
{gammaBrauer}).

These solutions give that in (\ref{K_alpha_0}),(\ref{K_kappa_1}),(\ref%
{K_kappa_2}),(\ref{K_alpha_2}),
\begin{eqnarray}
\alpha _{0} &=&2(d+e)+\frac{6(d-e)(r_{1}^{2}-r_{2}^{2})}{R^{2}},  \notag \\
\alpha _{2} &=&\alpha _{0}+c-4(d+e),  \notag \\
\kappa _{1} &=&\alpha _{0}-4d,  \notag \\
\kappa _{2} &=&\alpha _{0}-4e.  \label{solution_d_e_02}
\end{eqnarray}

Now we evaluate the $N_{12},M_{1},M_{2}$ in the large $r$ region, using the
general expressions (\ref{N_12}),(\ref{M_1}),(\ref{M_2}), as follows:%
\begin{eqnarray}
N_{12} &=&2K_{2}\partial _{t_{1}}\partial _{t_{2}}K_{0}-\partial
_{t_{1}}K_{1}\partial _{t_{2}}K_{1}+O(y^{2})  \notag \\
&=&\frac{r_{1}^{2}r_{2}^{2}}{a^{2}R^{8}}\frac{q^{3}}{aR^{2}}\left( \alpha
_{0}+\alpha _{2}-\kappa _{1}-\kappa _{2}\right) +O(1/r^{8})+O(y^{2})  \notag
\\
&=&\frac{q^{3}\mu _{1}^{2}\mu _{2}^{2}}{r^{6}}k+O(1/r^{8})+O(y^{2}),
\label{N_12_k}
\end{eqnarray}%
\begin{eqnarray}
M_{1} &=&-1+\frac{s_{2}\partial _{t_{1}}K_{1}-n_{12}\partial _{t_{2}}K_{1}}{%
s_{1}s_{2}-n_{12}^{2}}+O(y^{2})  \notag \\
&=&\frac{q}{aR^{2}}\left( -\alpha _{2}+\kappa _{1}\right)
+O(1/r^{4})+O(y^{2})  \notag \\
&=&-\frac{qm}{r^{2}}+O(1/r^{4})+O(y^{2}),  \label{M_1_m}
\end{eqnarray}%
and
\begin{eqnarray}
M_{2} &=&-1+\frac{s_{1}\partial _{t_{2}}K_{1}-n_{12}\partial _{t_{1}}K_{1}}{%
s_{1}s_{2}-n_{12}^{2}}+O(y^{2})  \notag \\
&=&\frac{q}{aR^{2}}\left( -\alpha _{2}+\kappa _{2}\right)
+O(1/r^{4})+O(y^{2})  \notag \\
&=&-\frac{qn}{r^{2}}+O(1/r^{4})+O(y^{2}),  \label{M_2_n}
\end{eqnarray}%
where we identify $r^{2}=aR^{2}(1+O(1/r^{2}))+O(y^{2})$ and $\frac{1}{r^{2}}=%
\frac{1}{aR^{2}}(1+O(1/r^{2}))+O(y^{2})$.\ We are making expansions in power
series of $y^{2}$ and$~1/r^{2}.$ More details of the above derivations are
in Appendix \ref{appendix_N12_M1_M2}.

In the expression of $M_{1}$, $M_{2}$, we have identified%
\begin{eqnarray}
q\left( \alpha _{2}-\kappa _{1}\right) &=&qm=q_{1},  \label{relation_m} \\
q\left( \alpha _{2}-\kappa _{2}\right) &=&qn=q_{2}.  \label{relation_n}
\end{eqnarray}

In the large $r$, we have the quantization of electric charges by%
\begin{eqnarray}
A^{i} &=&M_{i}dt, \\
F_{(2)}^{1} &=&dM_{1}dt=drdt\left( \frac{2q}{r^{3}}m+O(1/r^{5})\right) , \\
F_{(2)}^{2} &=&dM_{2}dt=drdt\left( \frac{2q}{r^{3}}n+O(1/r^{5})\right) ,
\end{eqnarray}%
\begin{eqnarray}
\frac{1}{4\pi ^{2}}\int \ast F_{(2)}^{1} &=&m, \\
\frac{1}{4\pi ^{2}}\int \ast F_{(2)}^{2} &=&n,
\end{eqnarray}%
where in our coordinates $g_{tt}g_{rr}=q$ in large $r$, and where $m,n$ are
the two charges. Note that terms corresponding to $O(1/r^{4})$ in (\ref%
{M_1_m}),(\ref{M_2_n}) will not contribute to the integral.

In the expression of $N_{12}$, we have identified
\begin{equation}
\alpha _{0}+\alpha _{2}-\kappa _{1}-\kappa _{2}=k.  \label{relation_k}
\end{equation}%
The two angles are the angles in the $z_{1}$ plane and $z_{2}$ plane
respectively.~

This $k$ will be identified with the $k$ parameter in the Brauer algebra
representation. \vspace{1pt}\vspace{1pt}In other words, for nonzero $k$,%
\begin{equation}
N_{12}=k\frac{\mu _{1}^{2}\mu _{2}^{2}q^{3}}{r^{6}}+O(1/r^{8}).
\end{equation}%
It appears at order $1/r^{6}$ with coefficients $k.$ For the special case $%
k=0,$ it may appear at only order $1/r^{8}.~$See Appendix \ref%
{appendix_N12_M1_M2} for more details. The difference between $M_{i}$ and $%
N_{i}~$is order $k/r^{4}$ for nonzero $k$, and $1/r^{6}$ for zero $k$.

Plugging these relations (\ref{relation_m}),(\ref{relation_n}),(\ref%
{relation_k}) \vspace{1pt}into the solutions (\ref{solution_d_e_02}), we find%
\begin{eqnarray}
4e &=&k-m=k-q_{1}/q,  \notag \\
4d &=&k-n=k-q_{2}/q,  \notag \\
c &=&k.
\end{eqnarray}

Note that in this family of solutions, the coefficients $e\leqslant
0,d\leqslant 0$ amount to that $k-m\leqslant 0,k-n\leqslant 0.$ Therefore
the sign property of $e,d$, which are either negative or zero, imposes the
constraints $k\leqslant m,n$, or equivalently $k\leqslant \min (m,n).$

In the $y$ expansions, we have, from (\ref{solution_d_e_02}),
\begin{eqnarray}
q\alpha _{0} &=&qk+3q_{1}\mu _{1}^{2}+3q_{2}\mu _{2}^{2}-2q_{1}-2q_{2}, \\
q\alpha _{2} &=&3q_{1}\mu _{1}^{2}+3q_{2}\mu _{2}^{2}-q_{1}-q_{2},
\label{parameter_alpha2_02} \\
q\kappa _{1} &=&3q_{1}\mu _{1}^{2}+3q_{2}\mu _{2}^{2}-2q_{1}-q_{2}, \\
q\kappa _{2} &=&3q_{1}\mu _{1}^{2}+3q_{2}\mu _{2}^{2}-q_{1}-2q_{2},
\label{parameters_01}
\end{eqnarray}%
where we used that $\mu _{1}^{2}+\mu _{2}^{2}=1-y^{2}/r^{2}=1-O(y^{2}).$

There are several equivalent and alternative ways of writing these
variables, see Appendix \ref{appendix_N12_M1_M2} for more details. For
example,
\begin{equation}
qk=q\alpha _{0}-q\alpha _{2}+q_{1}+q_{2}=q\alpha _{0}+q\alpha _{2}-q\kappa
_{1}-q\kappa _{2},
\end{equation}%
which will be frequently used in the derivations.

\vspace{1pt}

\section{Droplets and Young diagram operators}

\vspace{1pt}

\vspace{1pt}\renewcommand{\theequation}{7.\arabic{equation}} %
\setcounter{equation}{0}

\vspace{1pt}

\label{sec_mt_and_mapping}

\vspace{1pt}Now we turn to the analysis of the droplet configurations. The
solutions in section \ref{sec_N12_M1_M2_R_large} are the large $R$
expressions that result from the droplet configurations. The solution in
section \ref{sec_N12_M1_M2_R_large} are in the large $R$ region of the full
solution in all the ranges of $z_{i}$, $\bar{z}_{i}$. In this section, we
also provide further duality relation with the operators labelled by Brauer
algebra.

We have that from section \ref{sec_N12_M1_M2_R_large},
\begin{equation}
K_{0}=\frac{1}{2}aR^{2}-\frac{q}{2}\log (aR^{2})+\frac{q^{2}}{a}\left( \frac{%
(d+e)}{2R^{2}}+\frac{(d-e)(r_{1}^{2}-r_{2}^{2})}{2R^{4}}+O(1/r^{4})\right)
\label{K0_01}
\end{equation}%
in large $R$, and where%
\begin{eqnarray}
-4d &=&n-k,~~\ \ \ \ \  \\
-4e &=&m-k.\
\end{eqnarray}

As in section \ref{sec_flux_integration_01}, e.g. (\ref{flux_N_i}), the flux
quantization requires that the total droplet volume to be quantized,
\begin{equation}
\int_{\mathcal{D}}d^{2}z_{1}^{\prime }d^{2}z_{2}^{\prime }=\int_{\mathcal{D}%
(\emptyset )}d^{2}z_{1}^{\prime }d^{2}z_{2}^{\prime }=2\pi ^{3}l_{p}^{4}N
\end{equation}%
where $\mathcal{D}$ is the total $Z=-\frac{1}{2}$ droplets. Here we use the
notation that $d^{2}z_{1}d^{2}z_{2}=dx_{1}dx_{2}dx_{3}dx_{4}$ for
convenience. $\mathcal{D}(\emptyset )$ is the $Z=-\frac{1}{2}~$configuration
such that there is no any finite $Z=\frac{1}{2}~$domains or $Z=\frac{1}{2}$
bubbles. In the large $R$, it was shown \cite{Lin:2010nd} that, if we expand
\begin{equation}
K_{0}=\frac{1}{2}a(\left\vert z_{1}\right\vert ^{2}+\left\vert
z_{2}\right\vert ^{2})+\widetilde{K}_{0},\qquad K_{1}=\frac{1}{2}+\widetilde{%
K}_{1},
\end{equation}%
\begin{eqnarray}
&&\frac{1}{a}\partial _{1}\partial _{\bar{1}}\widetilde{K}_{0}+\partial
_{2}\partial _{\bar{2}}\widetilde{K}_{0}=\widetilde{K}_{1},
\label{K_tilde_eqn_01} \\
&&\partial _{1}\partial _{\bar{1}}\widetilde{K}_{1}+\partial _{2}\partial _{%
\bar{2}}\widetilde{K}_{1}=0.  \label{K_tilde_eqn_02}
\end{eqnarray}%
This means that a general solution to $\widetilde{K}_{0}$ is $-\frac{1}{2}%
\log (a\left\vert z_{1}-z_{1}^{\prime }\right\vert ^{2}+a\left\vert
z_{2}-z_{2}^{\prime }\right\vert ^{2})$ with ($z_{1}^{\prime }$,$\bar{z}%
_{1}^{\prime }$,$z_{2}^{\prime }$,$\bar{z}_{2}^{\prime }$) arbitrary,
therefore we can approximately expand $K_{0}$ as\vspace{1pt}%
\begin{eqnarray}
&&K_{0}=\frac{1}{2}aR^{2}-\frac{q}{2}\log (aR^{2})-\frac{q}{4\pi
^{3}l_{p}^{4}N}\int_{\mathcal{D}}d^{2}z_{1}^{\prime }d^{2}z_{2}^{\prime
}\log (a\left\vert z_{1}-z_{1}^{\prime }\right\vert ^{2}+a\left\vert
z_{2}-z_{2}^{\prime }\right\vert ^{2})  \notag \\
&&+\frac{q}{4\pi ^{3}l_{p}^{4}N}\int_{\mathcal{D}(\emptyset
)}d^{2}z_{1}^{\prime }d^{2}z_{2}^{\prime }\log (a\left\vert
z_{1}-z_{1}^{\prime }\right\vert ^{2}+a\left\vert z_{2}-z_{2}^{\prime
}\right\vert ^{2})+O(1/R^{4}).  \label{K0_02}
\end{eqnarray}%
Both (\ref{K0_01}),(\ref{K0_02}) satisfy the equations (\ref{K_tilde_eqn_01}%
),(\ref{K_tilde_eqn_02}). The droplet configuration in $\mathcal{D}$ gives
nontrivially the information of $d,e$ in the large $R$, when comparing two
expressions (\ref{K0_01}),(\ref{K0_02}).

Suppose we consider configurations that only depend on $r_{1},r_{2}$, and in
the case when $d$ and $e$ are equal, where there are more symmetry in the
droplet configuration, we can expand \vspace{1pt}(\ref{K0_02}),%
\begin{eqnarray}
K_{0} &=&\frac{1}{2}aR^{2}-\frac{q}{2}\log (aR^{2})-\frac{q(M_{_{11}}^{\left[
2\right] }-M_{_{11,\emptyset }}^{\left[ 2\right] })}{4\pi ^{3}l_{p}^{4}NR^{2}%
}-\frac{q(M_{_{22}}^{\left[ 2\right] }-M_{_{22,,\emptyset }}^{\left[ 2\right]
})}{4\pi ^{3}l_{p}^{4}NR^{2}}+O(1/R^{4})  \notag \\
&&
\end{eqnarray}%
where
\begin{eqnarray}
M_{_{11}}^{\left[ 2\right] }-M_{_{11,\emptyset }}^{\left[ 2\right]
} &=&\int_{\mathcal{D}}d^{2}z_{1}^{\prime }d^{2}z_{2}^{\prime
}\left\vert z_{1}^{\prime }\right\vert
^{2}-\int_{\mathcal{D}(\emptyset )}d^{2}z_{1}^{\prime
}d^{2}z_{2}^{\prime }\left\vert z_{1}^{\prime
}\right\vert ^{2}, \\
M_{_{22}}^{\left[ 2\right] }-M_{_{22,\emptyset }}^{\left[ 2\right]
} &=&\int_{\mathcal{D}}d^{2}z_{1}^{\prime }d^{2}z_{2}^{\prime
}\left\vert z_{2}^{\prime }\right\vert
^{2}-\int_{\mathcal{D}(\emptyset )}d^{2}z_{1}^{\prime
}d^{2}z_{2}^{\prime }\left\vert z_{2}^{\prime }\right\vert ^{2},
\end{eqnarray}%
are the second moments along the axis perpendicular to $z_{1}$ plane and $%
z_{2}$ plane respectively, subtracted from those of the configuration
without any finite $Z=\frac{1}{2}$ domains. The crossing terms $z_{i}\bar{z}%
_{i}^{\prime }$ cancels due to symmetric configuration in $z_{1}$ plane and
in $z_{2}$ plane.

We read off the coefficients that
\begin{eqnarray}
&&\frac{1}{2\pi ^{3}l_{p}^{4}N}(M_{_{11}}^{\left[ 2\right]
}-M_{_{11,\emptyset }}^{\left[ 2\right] })+\frac{1}{2\pi ^{3}l_{p}^{4}N}%
(M_{_{22}}^{\left[ 2\right] }-M_{_{22,\emptyset }}^{\left[ 2\right] }) \\
&=&\frac{1}{4}\frac{q}{a}[(n-k)+(m-k)]=-\frac{q}{a}(d+e).
\end{eqnarray}%
If we use the convention $\frac{q}{a}=r_{0}^{2}=(4\pi l_{p}^{4}N)^{1/2}$
e.g. as from (\ref{flux_a_02}),$~$and in the unit $r_{0}=1,$ \vspace{1pt}%
\begin{equation}
(M_{_{11}}^{\left[ 2\right] }-M_{_{11,\emptyset }}^{\left[ 2\right]
})+(M_{_{22}}^{\left[ 2\right] }-M_{_{22,\emptyset }}^{\left[ 2\right] })=%
\frac{\pi ^{2}}{8}[(n-k)+(m-k)].
\end{equation}%
We then identify the second moments in two directions,
\begin{eqnarray}
(M_{_{11}}^{\left[ 2\right] }-M_{_{11,\emptyset }}^{\left[ 2\right] }) &=&%
\frac{\pi ^{2}}{8}(n-k), \\
(M_{_{22}}^{\left[ 2\right] }-M_{_{22,\emptyset }}^{\left[ 2\right] }) &=&%
\frac{\pi ^{2}}{8}(m-k),
\end{eqnarray}%
where we are in the unit $r_{0}=1$.

Note that we also have the relation, from section \ref%
{sec_flux_integration_01},
\begin{eqnarray}
\sum_{i}\sum_{i^{\prime }=1}^{i}m_{i^{\prime }}N_{i} &=&m-k, \\
\sum_{i}\sum_{i^{\prime }=1}^{i}{\tilde{m}}_{i^{\prime }}\tilde{N}_{i}
&=&n-k.
\end{eqnarray}

The change in the potential $K_{0}$ is negative, when increasing the second
moments. It is analogous to the change of the potential from $-\log R^{2}$
to $-\log (R^{2}+\delta ^{2}$), due to increased second moments. So the
signs of $d,e$ are negative or zero in the above expression, that is
\begin{equation}
d\leqslant 0,\,\,\,\,e\leqslant 0.  \label{d_e_sign_property}
\end{equation}%
This sign property can be understood as that the potential in (\ref{K0_02})
will decrease when the $Z=-\frac{1}{2}$ droplets in the droplet space are
more outwards. The droplet second moments are increased from those of the
configuration when there is no any finite $Z=\frac{1}{2}~$domains. The
condition (\ref{d_e_sign_property}) is consistent with
\begin{equation}
\vspace{1pt}(M_{_{11}}^{\left[ 2\right] }-M_{_{11,\emptyset }}^{\left[ 2%
\right] })\geqslant 0,\,\,\,(M_{_{22}}^{\left[ 2\right] }-M_{_{22,\emptyset
}}^{\left[ 2\right] })\geqslant 0.  \label{condition_k_gravity_02}
\end{equation}

This amounts to
\begin{equation}
m-k\geqslant 0,\,\,\,\,n-k\geqslant 0
\end{equation}%
or equivalently
\begin{equation}
k\leqslant \min (m,n)  \label{condition_k_gravity_01}
\end{equation}%
from the gravity side.

The dual operator has dimension $m+n,$ and the two Young diagrams have $m-k$
and $n-k$ boxes respectively. The dimension is larger than the total number
of boxes by the amount $2k$. The $k$ parameter is identified in the last
section as in (\ref{N_12_k}),(\ref{relation_k}) as due to the mixing of two
angular directions. The two angles are the angles in the $z_{1}$ plane and $%
z_{2}$ plane respectively. We see that $2k$ measures the energy excess over
the total number of boxes, and is accounted for by the mixing of two angular
directions in the situation described here.

The total number of boxes match the calculation from the flux quantum
numbers, as $m-k$ and $n-k$ for two two-planes.

\begin{figure}[h]
\begin{center}
\includegraphics[width=14cm]{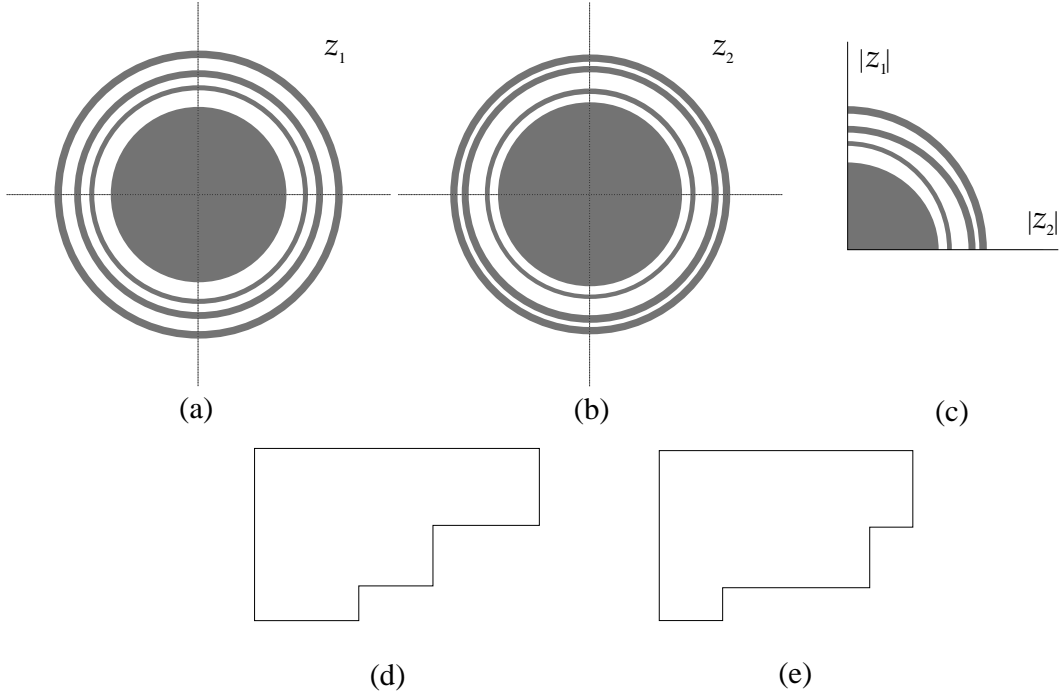}
\end{center}
\caption{{\protect\small Illustration of the mapping between Young diagrams
and the droplet configurations. $Z=-\frac{1}{2}$ are regions where $S^{3}$
vanishes and are drawn in black, while $Z=\frac{1}{2}$ are regions where $%
S^{1}$ vanishes and are drawn in white. In (c), there are black droplets and
white droplets in the $|z_{1}|,|z_{2}|$ quadrant. Figure (c) determines
(a,b), where (c) is projected onto $z_{1},z_{2}$ planes. The white regions
of (a),(b) map to the horizontal edges of (d),(e). The black regions of (c)
map to the vertical edges of (d),(e). The outward directions in the droplet
planes correspond to the upper-right directions along the edges of the Young
diagrams (d),(e). The flux quantum numbers in corresponding regions map to
the lengths of the edges of the Young diagrams (d),(e). This is an
illustration with the example of relatively few number of circles. More
general configurations involve many more circles in $z_{1},z_{2}$ planes.
The Young diagrams are filled with boxes which are not shown in the
illustration. }}
\label{young_5diagram01_01}
\end{figure}

The Brauer algebra provides two Young diagram representations, and
each correspond to droplet configurations in $z_{1}$ and $z_{2}$
planes respectively. For example, for configurations that have
$2l_{1}$ concentric circles in the droplet plane $z_{1}$, this
configuration maps to the $2l_{1}$ edges of the first Young
diagram $\gamma _{+}$, and the flux quantization numbers on each
droplet region map to the lengths of the edges of the Young
diagram $\gamma _{+}$. There are also $2l_{2}$ concentric
circles in the droplet plane $z_{2}$, and this configuration maps to the $%
2l_{2}$ edges of the second Young diagram $\gamma _{-}$, and the flux
quantization numbers on each droplet region map to the lengths of the edges
of the Young diagram $\gamma _{-}$. See figure \ref{young_5diagram01_01}.
These two Young diagrams are depicted in figures \ref{young_5diagram01_01}%
(d), \ref{young_5diagram01_01}(e) in the example.

\vspace{1pt}

We can characterize the droplet configuration by three diagrams,
for those depending only on $r_{1},r_{2}.$\ The first two diagrams
are concentric ring
patterns of alternative $Z=\frac{1}{2}$ and $Z=-\frac{1}{2}$ regions in $%
z_{1}$ planes and in $z_{2}$ planes, see figures \ref{young_5diagram01_01}%
(a), \ref{young_5diagram01_01}(b) for example. The third diagram is a
diagram in the $(r_{1},r_{2})$ quadrant, see figure \ref{young_5diagram01_01}%
(c).\ \vspace{1pt}We first map the corners in two Young diagrams to ordered
points along $r_{1},r_{2}$ axis respectively in figure \ref%
{young_5diagram01_01}(c). Then we draw diagrams connecting ordered points
along $r_{1}$ axis to those along $r_{2}~$axis, dividing regions in $%
(r_{1},r_{2})$ space into $Z=-\frac{1}{2}$ (depicted in black) and $Z=\frac{1%
}{2}$ (depicted in white).

We can also draw more complicated droplets. In the gravity description, when
we draw the lines, there are extra possibilities for possible lines to go in
the middle region of $(r_{1},r_{2})$ space. These may correspond to other
operators that are superpositions of the Brauer basis.

For those configurations that depend not only on $r_{1},r_{2}$,
they could be the superpositions of the Young diagram operators in
the above discussions. For example, one can add small ripples on
any boundaries of the droplets. See also related discussions, e.g.
\cite{Grant:2005qc}. The configurations that correspond to ripples
on the droplet boundaries, can be considered as the superpositions
of the Young diagram operators in the above discussions.

In \cite{Chen:2007du}, 1/2 BPS geometries were uplifted into the
systems of 1/4 and 1/8 BPS geometries, and there are disconnected
droplets with various topologies in 4d or 6d droplet space
respectively, and the topology change transitions occurred in
\cite{Lin:2004nb},\cite{Horava:2005pv} uplift to the topology
change transitions in 4d and 6d. In general, in the 4d droplet
space we study here, the topology change transition happens
commonly.

The condition (\ref{condition_k_gravity_01}) is consistent with
the range of $k$ in Brauer algebra representation. The droplet
information are encoded in the Young diagrams. There is
correspondence between the pair of Young diagrams and the
concentric droplet configuration in the two complex planes. The
operators labelled by the Young diagrams of Brauer algebra give a
family of globally well-defined spacetime geometries. Other bases
may also be related to the droplet picture, since they can be
related by transformations, and it might be interesting to see how
$k$ is produced in other bases. The system of geometries are dual
to the system of the corresponding operators.

\vspace{1pt}

\section{Gauge invariant operators by Brauer algebra}

\vspace{1pt}\renewcommand{\theequation}{8.\arabic{equation}} %
\setcounter{equation}{0}

\vspace{1pt}

\vspace{1pt}

\label{sec_operater_side}


Now we will briefly summarize the operators based on the Brauer algebra \cite%
{Kimura:2007wy},\cite{Kimura:2008ac}. See also related discussion in \cite%
{Kimura:2009jf},\cite{Kimura:2009ur}.

We take two complex fields $X$, $Y$ out of three complex fields and consider
gauge invariant operators constructed from $m$ $X$s and $n$ $Y$s. The $U(N)$
gauge group is considered. The operators are conveniently expressed in the
notation of \cite{Kimura:2007wy},\cite{Kimura:2008ac} by
\begin{equation}
O_{A,ij}^{\gamma }(X,Y)=tr_{m,n}\left( Q_{A,ij}^{\gamma }X^{\otimes
m}\otimes (Y^{T})^{\otimes n}\right) ,  \label{brauerbasis}
\end{equation}%
where $Q_{A,ij}^{\gamma }$ is given by a linear combination of elements in
the Brauer algebra. Irreducible representations of the Brauer algebra are
denoted by $\gamma $, which are given by two Young diagrams:
\begin{equation}
\gamma =(\gamma _{+},\gamma _{-}),  \label{gammaBrauer}
\end{equation}%
where $\gamma _{+}$ is a Young diagram with $m-k$ boxes and $\gamma _{-}$ is
a Young diagram with $n-k$ boxes. $k$ is an integer satisfying $0\leqslant
k\leqslant \min (m,n)$. $A=(R,S)$ is an irreducible representation of $%
S_{m}\times S_{n}$, labelled by two Young diagrams with $m$ and $n$ boxes. $%
i,j$ are multiplicity indices with respect to the embedding of $A$ into $%
\gamma $.

An advantage of this basis is that the free two-point functions are
diagonal. We also note that non-planar corrections are fully taken into
account.

Because the number of boxes in $\gamma $ is characterized by $k$, it is
convenient to classify the operators by the integer $k$.

The labels can be simplified when $k=0$, because $i,j$ are trivial, and we
have $\gamma _{+}=R$, $\gamma _{-}=S$. Hence the operators in $k=0$ are
labelled by two Young diagrams. Denoting $P_{R,S}=Q_{A,ij}^{\gamma (k=0)}$,
the expression in (\ref{brauerbasis}) becomes
\begin{equation}
O_{R,S}(X,Y)=tr_{m,n}\left( P_{R,S}X^{\otimes m}\otimes (Y^{T})^{\otimes
n}\right) .  \label{brauerk=0}
\end{equation}%
In this sector, the operators have the nice expansion with respect to $1/N$,
whose leading term is given by
\begin{equation}
O_{R,S}(X,Y)=O_{R}(X)O_{S}(Y)+\cdots .  \label{k=0leading}
\end{equation}%
Here $O_{R}(X)$ is the Schur polynomial built from the $X$ fields labelled
by a Young diagram $R$ with $m$ boxes, and $O_{S}(Y)$ is the Schur
polynomial built from the $Y$ fields labelled by a Young diagram $S$ with $n$
boxes.

In general, including the case $k=0$, the leading term of the operators (\ref%
{brauerbasis}) looks schematically like
\begin{equation}
O_{A,ij}^{\gamma }(X,Y)\sim tr_{m,n}\left( \sigma C^{k}X^{\otimes m}\otimes
(Y^{T})^{\otimes n}\right) +\cdots ,  \label{contraction}
\end{equation}%
where $\sigma $ is an element in $S_{m}\times S_{n}$ and $C$ is an operation
contracting the upper index of an $X$ and the upper index of a $Y^{T}$. Each
term in$~$the dots in the above expression (\ref{contraction}) contains more
contractions. In other words, $k$ is the minimum number of contractions
involved in the $Q_{A,ij}^{\gamma }$ of an operator. Therefore the $k$ can
be given the intuitive meaning that it measures the degree of the mixing
between the two fields. For example, the $k=0$ has no mixing in the sense of
(\ref{k=0leading}). The opposite case is the case $k$ takes the maximum
value. With the condition $m=n$, operators in $k=m=n$ are found to be
expressed by 
operators of the combined matrix $XY$ \cite{Kimura:2009ur}.

In the paper \cite{Kimura:2010tx}, a class of the $1/4$ BPS operators were
constructed by exploiting algebraic properties of the Brauer algebra, where
it was shown that the operators $O_{R,S}(X,Y)$ and $\sum_{R,S,i}O_{A,ii}^{%
\gamma \hspace{0.05cm}(k\neq 0)}(X,Y)$ are annihilated by the one-loop
dilatation operator, for any $m$, $n$ and $N$. Defining $P^{\gamma
}=\sum_{A,i}Q_{A,ii}^{\gamma }$ for $k\neq 0$, the BPS operators are
presented by
\begin{equation}
O^{\gamma }(X,Y)=tr_{m,n}(P^{\gamma }X^{\otimes m}\otimes (Y^{T})^{\otimes
n})  \label{brauerBPS}
\end{equation}%
for both $k=0$ and $k\neq 0$. Note that $P^{\gamma}$ is the projector
associated with an irreducible representation $\gamma$ of the Brauer algebra.

It is interesting to find that they are labelled by two Young diagrams with
boxes whose total number depends on the integer $k$ for fixed $m$ and $n$.
In $k=0$, the number of boxes involved in a representation $\gamma $ is
equal to the sum of the R-charges. On the other hand, when $k$ is non-zero,
the Young diagrams $\gamma =(\gamma _{+},\gamma _{-})$ have a smaller number
of boxes than the sum of the R-charges. The number of the deficit boxes in
the $\gamma $ is $2k$.

When $k$ is equal to $n$ for $m\geqslant n$, the operators are labelled by a
single Young diagram with $m-k$ boxes. A special case happens for $k=m$ with
$m=n$. The operator does not have any Young diagrams, but this is different
from the case $m=n=0$. For a given $m=n$, there is only one 1/4 BPS operator
labelled by the trivial representation.

We now have a remark on a constraint for representations of the Brauer
algebra. The representations $\gamma $ have the constraint $c_{1}(\gamma
_{+})+c_{1}(\gamma _{-})\leq N$, where $c_{1}$ denotes the length of the
first column of the Young diagram. More explanations are provided in \cite%
{Kimura:2007wy}. This constraint is consistent with the identifications in
sections \ref{sec_flux_integration_01} and \ref{sec_mt_and_mapping}.

The Brauer basis can be used in various ways. One of the motivations of
constructing the Brauer basis in \cite{Kimura:2007wy} was to construct an
operator describing a set of D-branes and anti-D-branes, which is realized
as the $k=0$ sector. In the application of the Brauer basis to the $su(2)$
sector, which is relevant for the present work, the $k=0$ sector realizes
natural operators dual to the objects with two fields. Furthermore the
construction of the basis introduces the operators labelled by $k\neq 0$ as
well. In the mapping proposed in this paper, the quantum number $k$ has been
given a meaning as the mixing between the two angular directions from the
gravity point of view. The use of the Brauer algebra may be rephrased as the
manifestation of such a good quantum number.

On the other hand, the BPS operators may be described by other bases
diagonalizing free two-point functions, as in \cite{Brown:2007xh},\cite%
{Bhattacharyya:2008rb}, see also related discussions, \cite{Carlson:2011hy},%
\cite{Brown:2010pb},\cite{Pasukonis:2010rv},\cite{DeComarmond:2010ie}, \cite%
{Koch:2011hb}. Because other bases respect other quantum numbers,
understanding a map between the two sides for other bases could also be
useful for getting a complete duality of this sector.


\vspace{1pt} \vspace{1pt}

\section{Asymptotics of metric}

\vspace{1pt}

\vspace{1pt}\renewcommand{\theequation}{9.\arabic{equation}} %
\setcounter{equation}{0}

\vspace{1pt}

\label{sec_geometry_large_r}


In this section we analyze the large $r$ asymptotics of the geometry. We
start from the expression (\ref{geometries_new_variable_01}). The large $r$
region includes the large $R$ region of the droplet space. We first expand
in small $y$, and have $K_{0},K_{1},K_{2}$. We then expand these functions
in powers of $1/r^{2}$. In the large $r$, we have
\begin{equation}
\frac{r_{1}^{2}}{R^{2}}=\mu _{1}^{2}\left( 1+O(1/r^{2})\right) ,~~~~\frac{%
r_{2}^{2}}{R^{2}}=\mu _{2}^{2}\left( 1+O(1/r^{2})\right) ,
\end{equation}%
\begin{equation}
R_{1}=\frac{1}{\sqrt{a}}\sqrt{r^{2}+qC_{1}}\left( 1+O(1/r^{2})\right) ,\quad
R_{2}=\frac{1}{\sqrt{a}}\sqrt{r^{2}+qC_{2}}\left( 1+O(1/r^{2})\right) .
\end{equation}

Near $y=0$, but large $R$, from (\ref{deltaexpansion}),%
\begin{eqnarray}
\Delta &=&\frac{1}{4r^{4}K_{2}}+O(y^{2})  \notag \\
&=&\frac{a^{2}R^{4}}{r^{4}q}\left( 1-\frac{q(2+\alpha _{2})}{aR^{2}}%
+O(1/r^{4})\right) +O(y^{2})  \notag \\
&=&\frac{1}{q}\left( 1+\frac{q(2C_{i}\mu _{i}^{2}-2-\alpha _{2})}{r^{2}}%
\right) +O(1/r^{4})+O(y^{2})  \notag \\
&=&\frac{1}{q}\left( 1+\frac{q(q_{1}+q_{2}-q_{1}\mu _{1}^{2}-q_{2}\mu
_{2}^{2})}{r^{2}}\right) +O(1/r^{4})+O(y^{2})
\end{eqnarray}%
where we have used (\ref{parameter_alpha2_02}) and have identified
\begin{equation}
qC_{1}=q+q_{1},\quad qC_{2}=q+q_{2}.  \label{Ci_q}
\end{equation}%
We find that $\Delta $ has the same asymptotic form as the two-charge
superstar.

The asymptotic form of the metric can be rewritten in the following form
\begin{equation}
ds^{2}=\sqrt{\Delta }ds_{1}^{2}+\frac{1}{\sqrt{\Delta }}ds_{2}^{2}
\end{equation}%
analogous to the gauged supergravity ansatz. We have that
\begin{eqnarray}
ds_{1}^{2} &=&-\left( r^{2}+q-\frac{q^{2}(C_{1}\mu _{1}^{2}+C_{2}\mu
_{2}^{2}-1+\alpha _{2}-2\kappa _{1}\mu _{1}^{2}-2\kappa _{2}\mu _{2}^{2})}{%
r^{2}}+O(1/r^{4})\right) dt^{2}  \notag \\
&&+\frac{q}{r^{2}}\left( 1-\frac{q(3C_{1}\mu _{1}^{2}+3C_{2}\mu
_{2}^{2}-\alpha _{2}-2)}{r^{2}}+O(1/r^{4})\right) dr^{2}+r^{2}d\Omega
_{3}^{2}  \label{asymptoticmetric1}
\end{eqnarray}%
and
\begin{eqnarray}
ds_{2}^{2} &=&d\mu _{3}^{2}+\mu _{3}^{2}d\psi ^{2}+H_{1}\left[ d\mu
_{1}^{2}+\mu _{1}^{2}\left( 1+O(1/r^{4})\right) (d\phi _{1}+M_{\phi
_{1}}dt)^{2}\right]  \notag  \label{asymptoticmetric2} \\
&&+H_{2}\left[ d\mu _{2}^{2}+\mu _{2}^{2}\left( 1+O(1/r^{4})\right) (d\phi
_{2}+M_{\phi _{2}}dt)^{2}\right]  \notag \\
&&+O(1/r^{5})d\mu _{1}dr+O(1/r^{5})d\mu _{2}dr  \notag \\
&&+2\frac{q^{2}}{r^{4}}(C_{1}+C_{2}+\alpha _{0}-\alpha
_{2}-2)(1+O(1/r^{2}))\mu _{1}\mu _{2}d\mu _{1}d\mu _{2}  \notag \\
&&+2\frac{q^{2}\mu _{1}^{2}\mu _{2}^{2}}{r^{4}}(\alpha _{2}+\alpha
_{0}-\kappa _{1}-\kappa _{2})(1+O(1/r^{2}))(d\phi _{1}+M_{\phi
_{1}}dt)(d\phi _{2}+M_{\phi _{2}}dt)  \notag \\
&&
\end{eqnarray}%
where the $M_{i}$ were calculated in the section \ref{sec_N12_M1_M2_R_large}%
,
\begin{eqnarray}
&&M_{1}=\frac{q}{r^{2}}(-\alpha _{2}+\kappa _{1})+O(1/r^{4})+O(y^{2}), \\
&&M_{2}=\frac{q}{r^{2}}(-\alpha _{2}+\kappa _{2})+O(1/r^{4})+O(y^{2}).
\end{eqnarray}%
See Appendix \ref{appendix_derivation_metric_r_large} for detailed
derivation.

When we use (see section \ref{sec_N12_M1_M2_R_large} and Appendix \ref%
{appendix_N12_M1_M2}, e.g. equation (\ref{parameter_alpha2_02})),%
\begin{eqnarray}
q\kappa _{1} &=&q\alpha _{2}-q_{1}, \\
q\kappa _{2} &=&q\alpha _{2}-q_{2}, \\
q\alpha _{2} &=&3q_{1}\mu _{1}^{2}+3q_{2}\mu _{2}^{2}-q_{1}-q_{2},
\label{alpha2bymuq}
\end{eqnarray}%
for the $g_{tt}$ and $g_{rr}$, we get
\begin{eqnarray}
ds_{1}^{2} &=&-\left( r^{2}+1-\frac{q_{1}+q_{2}}{r^{2}}+O(1/r^{4})\right)
dt^{2}  \notag \\
&&+\frac{1}{r^{2}}\left( 1-\frac{q_{1}+q_{2}+1}{r^{2}}+O(1/r^{4})\right)
dr^{2}+r^{2}d\Omega _{3}^{2}
\end{eqnarray}%
and we have rescaled $r^{2}\rightarrow qr^{2}$. By a rescaling $%
r^{2}\rightarrow qr^{2}$, $q$ appears as an overall factor of the metric.
The metric has the same asymptotic form as two-charge superstar \cite%
{Behrndt:1998ns},\cite{Myers:2001aq}.

When we use
\begin{equation}
qk=q\alpha _{0}-q\alpha _{2}+q_{1}+q_{2}=q\alpha _{0}+q\alpha _{2}-q\kappa
_{1}-q\kappa _{2},
\end{equation}%
then the last two terms in $ds_{2}^{2}$ becomes
\begin{equation}
2\frac{\mu _{1}\mu _{2}}{r^{4}}kd\mu _{1}d\mu _{2}+2\frac{\mu _{1}^{2}\mu
_{2}^{2}}{r^{4}}k(d\phi _{1}+M_{\phi _{1}}dt)(d\phi _{2}+M_{\phi _{2}}dt),
\label{angles_mixing}
\end{equation}%
and we have rescaled $r^{2}\rightarrow qr^{2}$.

The asymptotic geometry is similar in form to the \vspace{1pt}$U(1)^{3}~$%
gauged supergravity ansatz. One of the differences is that we have
additional mixing of $\mu _{1},\mu _{2}$, and mixing of $\phi _{1},\phi
_{2}, $ as in (\ref{asymptoticmetric2}) or (\ref{angles_mixing}). These
mixing terms correspond to the $k$ parameter in Brauer algebra.

\section{Discussions}

\renewcommand{\theequation}{9.\arabic{equation}} \setcounter{equation}{0}

\label{discussion}

We studied the characterization of the droplet configurations of the 1/4 BPS
geometries which are dual to a family of 1/4 BPS operators with large
dimensions in $\mathcal{N}$=$4$ SYM. We characterized the 4d droplet
configurations underlying the 1/4 BPS geometries, following early works. The
droplet space is enlarged from the 2d droplet space observed in the 1/2 BPS
case. The droplet regions are divided into two regions, and we projected the
droplet configuration into two complex planes. We map the concentric circle
patterns in the $z_{1}$ plane and $z_{2}$ plane to two Young diagrams. We
identify these two Young diagrams as the two Young diagrams in the operators
of the Brauer basis \cite{Kimura:2007wy},\cite{Kimura:2010tx}, which have
total number of boxes $m-k$ and $n-k$. The flux quantum numbers on the
droplets map to the edges of the Young diagrams and the radial directions in
the two-planes correspond to the upper-right directions along the edges of
the two Young diagrams.

We simplified the droplet configurations by projecting it to two
two-planes, and draw three diagrams. The first two diagrams are
black/white coloring on
the two-planes, and the third diagram is the black/white 
coloring on the ($r_{1},r_{2}$) space. We simplified droplet
configurations in particular by the third diagram. These include
general radially symmetric configurations in two two-planes.

There is also a droplet description from other method on the gauge side by
\cite{Berenstein:2005aa}. We see more consistency suggesting it to be the 4d
droplet space related to the multi-body system. There are some subtleties in
this droplet space.

We studied more about the small $y$ expansion of the geometries. In
particular, $K_{0}$ can be viewed as a potential on the droplet space, and
itself is determined by the total droplet configurations. We also performed
the large $R$ analysis in the droplet space, and find their connections with
the large $r$ asymptotics of the geometries. The large $R$ in the droplet
space encodes information of the large $r$ asymptotics. These information
are encoded in the $K_{0},K_{1},K_{2}$ in the small $y$ expansion. The large
$R$ expansion captures the large $r$ expansion. The asymptotics knows $m,n,k$%
, since $J_{1}=(m-k)+k$, $J_{2}=(n-k)+k$. The droplet configuration captures
$m-k,n-k,k$ and the details of all the information of the shapes of two
Young diagrams.

We identified families of geometries that have mixing between two angular
directions which are the two angles $\phi _{1},\phi _{2}$ in the $z_{1}$
plane and $z_{2}$ plane. The geometries in the asymptotic region have mixing
metric-component in $h_{\phi _{1}\phi _{2}}$ with a family of parameter $k$.
An interesting observation was given that the parameter $k$ have to satisfy $%
k\leqslant \min (m,n)$. We gave the interpretation that this
parameter is identified with the $k$ parameter in the Brauer
algebra representations.

Performing a similar analysis for the 1/8 BPS sector would raise an
interesting question. A proper basis using Brauer algebras has not been
constructed to deal with 1/8 BPS operators. One may guess that three
integers would be involved as the coefficients of the mixing among the three
angular directions, generalizing the one integer $k$ in the present case.
Such analysis may give rise to a hint to apply Brauer algebra for gauge
invariant operators involving more kinds of fields than two.

We mainly provided a mapping for the operators built from the projector of
the Brauer algebra in (\ref{brauerBPS}). However, this above-mentioned class
of expressions do not cover all types of 1/4 BPS operators. In the droplet
picture, there are also other more complicated droplet configurations. It
would be nice to understand other types of BPS operators from the Brauer
algebra, as well as in other bases. The droplet configurations on gravity
side would be helpful to get a complete list of the BPS operators
manipulated by the Brauer algebra.

\vspace{0.04cm} We can also study other excitations on these
states. One can consider the supergravity field excitations on
them. We can also consider strings excited on them. One can also
see the emergence of the other local excitations on the geometries, e.g. \cite{Chen:2007gh},\cite%
{Koch:2008ah},\cite{Lin:2010sba}.

\vspace{0.04cm} We may view the non-BPS states as the excitation
above the BPS states. Starting from these heavy BPS states, one
can add additional non-BPS excitations on them. These can be done
by modifying the operators by adding other fields or multiplying
other fields. These studies will also provide another view on the
physical meaning of the parameters of the bases.

\section*{Acknowledgments}

\vspace{1pt}

The work of Y.K. is supported in part by the research grants MICINN
FPA2009-07122 and MEC-DGI CSD2007-00042. We thank Jonathan P. Shock for
collaboration at an early stage. The work of H.L. is supported in part by
Xunta de Galicia (Conselleria de Educacion and grants PGIDIT10PXIB 206075PR
and INCITE09 206121PR), by the Spanish Consolider-Ingenio 2010 Programme
CPAN (CSD2007-00042), by the Juan de la Cierva of MICINN, and by the ME,
MICINN and Feder (grant FPA2008-01838). We also would like to thank
University of Valencia for hospitality. H. Lin also would like to thank the
UNIFY International Workshop on Frontiers in Theoretical Physics, University
of Porto, and University of Cambridge for hospitalities.

\appendix

\renewcommand{\theequation}{A.\arabic{equation}} \setcounter{equation}{0}

\vspace{1pt}

\vspace{1pt}

\section{Review of gravity ansatz}

\vspace{1pt}\renewcommand{\theequation}{A.\arabic{equation}} %
\setcounter{equation}{0}

\vspace{1pt}\label{appendix_gravity_a}

We review the geometry and the ansatz for the 1/4 BPS configurations. These
backgrounds have an additional $S^{1}$ isometry compared with the 1/8 BPS
geometries, and have a ten-dimensional solution of the form, in the
conventions of \cite{Donos:2006iy},\cite{Chen:2007du},\cite{Lunin:2008tf},
\begin{eqnarray}
ds_{{10}}^{2} &=&-h^{-2}(dt+\omega )^{2}+h^{2}((Z+\frac{1}{2})^{{-1}%
}2\partial _{i}\bar{\partial}_{j}Kdz^{i}d{\bar{z}}^{\bar{j}%
}+dy^{2})+y(e^{G}d\Omega _{3}^{2}+e{^{-G}}(d\psi +{\mathcal{A}})^{2}),
\notag \\
F_{5} &=&\left\{ -d{[}y^{2}e^{{2G}}(dt+\omega ){]}-y^{2}(d\omega +\eta
\mathcal{F})+2i\partial \bar{\partial}K\right\} \wedge d\Omega ^{3}+\mathrm{%
dual,}  \notag \\
h^{{-2}} &=&2y\cosh G,\qquad  \notag \\
Z &=&\frac{1}{2}\mathrm{\tanh }G=-\frac{1}{2}y\partial _{y}(\frac{1}{y}%
\partial _{y}K),\,  \notag \\
d\omega &=&\frac{i}{2}d[\frac{1}{y}\partial _{{y}}(\bar{\partial}-\partial
)K]=\frac{i}{y}(\partial _{i}\partial _{{\bar{j}}}\partial _{{y}}Kdz_{i}d%
\bar{z}_{\bar{j}}+\partial _{\bar{{i}}}Zd\bar{z}_{i}dy-\partial _{{i}%
}Zdz_{i}dy),  \notag \\
2\eta \mathcal{F} &=&-i\partial \bar{\partial}D.  \label{1-4_ansatz}
\end{eqnarray}%
$K=K(z_{i},\bar{z}_{i};y),$ where $i=1,2,~$is the K\"{a}hler potential for
the 4d base, which also varies with the $y$ direction. $D$ can be set to a
constant, if the fibration of the $S^{1}$\ is a direct product.~The volume
of the 4d base is constrained by a Monge-Ampere equation, as well as an
equation for function $D$,
\begin{eqnarray}
&&\log \det h_{i\bar{j}}=\log (Z+\frac{1}{2})+n\eta \log y+\frac{1}{y}%
(2-n\eta )\partial _{y}K+D(z_{i},\bar{z}_{\bar{j}}),  \label{eqn_general} \\
&&(1+\ast _{4})\partial \bar{\partial}D=\frac{4}{y^{2}}(1-n\eta )\partial
\bar{\partial}K.  \label{eq:MA2}
\end{eqnarray}%
In other words,
\begin{equation}
\det \partial _{i}\partial _{\bar{j}}K=(Z+\frac{1}{2})y^{n\eta }e^{\frac{1}{y%
}(2-n\eta )\partial _{y}K}e^{D}.  \label{MA_general}
\end{equation}%
\vspace{1pt}According to the analysis of \cite{Chen:2007du},\cite%
{Lunin:2008tf},\cite{Lin:2010nd} a family of geometries have the expansion
from the droplet space as:
\begin{equation}
K=-\frac{1}{4}y^{2}\log (y^{2})+K_{0}(z_{i},\bar{z}_{i})+y^{2}K_{1}(z_{i},%
\bar{z}_{i})+(y^{2})^{2}K_{2}(z_{i},\bar{z}_{i})+\sum_{n\geqslant
3}(y^{2})^{n}K_{n}(z_{i},\bar{z}_{i})  \label{K_01_02}
\end{equation}%
for $Z=\frac{1}{2}$ and%
\begin{equation}
K=\frac{1}{4}y^{2}\log (y^{2})+K_{0}(z_{i},\bar{z}_{i})+y^{2}K_{1}(z_{i},%
\bar{z}_{i})+(y^{2})^{2}K_{2}(z_{i},\bar{z}_{i})+\sum_{n\geqslant
3}(y^{2})^{n}K_{n}(z_{i},\bar{z}_{i})  \label{K_01_01}
\end{equation}%
for $Z=-\frac{1}{2}$,$~$with $\partial _{i}\partial _{\bar{j}}K_{0}=0$. The
last terms above correspond to expansion with higher order terms $%
(y^{2})^{n}K_{n}(z_{i},\bar{z}_{i})$. The $K_{0}(z_{i},\bar{z}_{i})$ is a
function defined on the droplet space.

As was shown in \cite{Lin:2010nd}, all other higher $K_{n}$, with $%
n\geqslant 3$, are expressed in terms of $K_{0},K_{1},K_{2}$, and thus the
entire solutions to $K$ are given uniquely by $K_{0},K_{1},K_{2}.$

$K_{0},K_{1}$ are determined by the coupled equations on the $Z=\frac{1}{2}$
droplet region \cite{Lin:2010nd}: 
\begin{eqnarray}
&&(\partial _{1}\partial _{{\bar{1}}}K_{0})(\partial _{2}\partial _{\bar{2}%
}K_{1})+(\partial _{1}\partial _{\bar{1}}K_{1})(\partial _{2}\partial _{\bar{%
2}}K_{0})-2(\partial _{1}\partial _{\bar{2}}K_{0})(\partial _{\bar{1}%
}\partial _{2}K_{1})=0,  \notag \\
&&\det \partial _{i}\partial _{\bar{j}}K_{0}=\frac{1}{4e}e^{2K_{1}}.
\label{coupledeqK0K1}
\end{eqnarray}

For the $AdS_{5}\times S^{5}~$,
\begin{equation}
K_{0}=\left\{
\begin{array}{c}
\frac{1}{2}aR^{2}-\frac{1}{2}ar_{0}^{2}\mathrm{\log (}%
R^{2}/r_{0}^{2}),~~~~~~~R^{2}\geqslant r_{0}^{2} \\
\frac{1}{2}ar_{0}^{2},~\ \ ~\ \ \ \ \ \ ~~\ ~~~\ \ \ ~~~\ ~~\ 0\leqslant
R^{2}\leqslant r_{0}^{2}%
\end{array}%
\right.  \label{K_0_ads}
\end{equation}%
as analyzed in \cite{Lin:2010nd}, where $Z=-\frac{1}{2}$ droplets are in $%
0\leqslant R^{2}\leqslant r_{0}^{2}$ and $Z=\frac{1}{2}$ droplets are in $%
~R^{2}\geqslant r_{0}^{2}.$ $~$Here, the $K_{0}$ is constant in $Z=-\frac{1}{%
2}$ droplets. In the above, $R^{2}=\left\vert z_{1}\right\vert
^{2}+\left\vert z_{2}\right\vert ^{2}$, in this appendix. The plot in figure %
\ref{plot_01_01} is a more general situation when there are also other three
$Z=-\frac{1}{2}$ strips; and in the limit when these three strips go to
zero, it recovers the expression (\ref{K_0_ads}). One can also introduce an
overall constant shift in $K_{0}$, since only its derivatives appear. One
can also rescale the above expression by the rescaling transformations as in
section \ref{sec_N12_M1_M2_R_large}.

\section{Derivation of the metric}

\vspace{1pt}\renewcommand{\theequation}{B.\arabic{equation}} %
\setcounter{equation}{0}

\vspace{1pt}

\label{appendix_derivation_metric}

\vspace{1pt}

In this appendix we derive the general form of the metric, using new
variables in the section \ref{sec_metric_01}.

For the situation that the K\"{a}hler potential does not depend on
the angular coordinates $\phi _{1}$, $\phi _{2}$, that
$K=K(r_{1},r_{2},y)$, the metric (\ref{originalmetric}) can be
expressed by
\begin{eqnarray}
ds^{2} &=&-h^{-2}\left( dt^{2}+2(\omega _{\phi _{1}}d\phi _{1}+\omega _{\phi
_{2}}d\phi _{2})dt+\omega _{\phi _{1}}^{2}d\phi _{1}^{2}+\omega _{\phi
_{2}}^{2}d\phi _{2}^{2}+2\omega _{\phi _{1}}\omega _{\phi _{2}}d\phi
_{1}d\phi _{2}\right)  \notag \\
&&+h^{2}\left( \mu _{3}^{2}dr^{2}+r^{2}d\mu _{3}^{2}+S_{11}R_{1}^{2}(d\mu
_{1}^{2}+\mu _{1}^{2}d\phi _{1}^{2})+S_{22}R_{2}^{2}(d\mu _{2}^{2}+\mu
_{2}^{2}d\phi _{2}^{2})\right.  \notag \\
&&\left. +\left( S_{11}\mu _{1}^{2}T_{1}^{2}+S_{22}\mu
_{2}^{2}T_{2}^{2}+2S_{12}\mu _{1}\mu _{2}T_{1}T_{2}\right) dr^{2}\right.
\notag \\
&&\left. +2\left( S_{11}R_{1}T_{1}\mu _{1}+S_{12}T_{2}\mu _{2}R_{1}-\mu
_{1}r\right) d\mu _{1}dr\right.  \notag \\
&&\left. +2\left( S_{22}R_{2}T_{2}\mu _{2}+S_{12}T_{1}\mu _{1}R_{2}-\mu
_{2}r\right) d\mu _{2}dr\right.  \notag \\
&&\left. +2S_{12}R_{1}R_{2}d\mu _{1}d\mu _{2}+2S_{12}R_{1}R_{2}\mu _{1}\mu
_{2}d\phi _{1}d\phi _{2}\right)  \notag \\
&&+ye^{G}d\Omega _{3}^{2}+ye^{-G}d\psi ^{2},
\label{NewCoordinateMetricBeforeShift}
\end{eqnarray}%
where we have used $\omega =\omega _{\phi _{1}}d\phi _{1}+\omega _{\phi
_{2}}d\phi _{2}$ and $S_{ij}=S_{ji}$, which are the consequence of the
assumption. We have defined $T_{i}=dR_{i}/dr$.

Performing the shift of the angular variables $\phi _{i}\rightarrow \phi
_{i}-t$, $i=1,2$, the metric can be further written to be the form
\begin{eqnarray}
ds^{2} &=&-h^{-2}\left( 1+h_{ab}M_{a}M_{b}-S_{t}\right) dt^{2}+h^{2}\left(
\mu _{3}^{2}+S_{11}\mu _{1}^{2}T_{1}^{2}+S_{22}\mu
_{2}^{2}T_{2}^{2}+2S_{12}\mu _{1}\mu _{2}T_{1}T_{2}\right) dr^{2}  \notag \\
&+&2h^{2}(S_{11}R_{1}T_{1}\mu _{1}+S_{12}T_{2}\mu _{2}R_{1}-\mu _{1}r)d\mu
_{1}dr+2h^{2}(S_{22}R_{2}T_{2}\mu _{2}+S_{12}T_{1}\mu _{1}R_{2}-\mu
_{2}r)d\mu _{2}dr  \notag \\
&+&\sqrt{\Delta }r^{2}d\Omega _{3}^{2}+\frac{1}{\sqrt{\Delta }}\left( d\mu
_{3}^{2}+H_{1}d\mu _{1}^{2}+H_{2}d\mu _{2}^{2}\right) +2h^{2}\left(
S_{12}R_{1}R_{2}-\frac{\mu _{1}\mu _{2}}{\Delta }\right) d\mu _{1}d\mu _{2}
\notag \\
&+&\frac{\mu _{3}^{2}}{\sqrt{\Delta }}d\psi ^{2}+h^{-2}h_{ab}(d\phi
_{a}+M_{a}dt)(d\phi _{b}+M_{b}dt).
\end{eqnarray}

The more detailed computations are given below.

%

\subsection{Metric functions}

We introduce $\Delta $ by the equation
\begin{equation}
e^{G}=\frac{r\sqrt{\Delta }}{\mu _{3}}.
\end{equation}%
Then $Z$ and $h^{-2}$ may be expressed as
\begin{eqnarray}
&&Z=\frac{1}{2}\tanh G=\frac{1}{2}\frac{r^{2}\Delta -\mu _{3}^{2}}{%
r^{2}\Delta +\mu _{3}^{2}},  \label{defZbyDelta} \\
&&h^{-2}=2y\cosh G=\frac{r^{2}\Delta +\mu _{3}^{2}}{\sqrt{\Delta }}.
\end{eqnarray}%
(\ref{defZbyDelta}) may be used to give $\Delta $ in terms of $Z$ as
\begin{equation}
\Delta =\frac{\mu _{3}^{2}}{r^{2}}\frac{1+2Z}{1-2Z}.
\end{equation}

\subsection{Terms with $d \protect\mu_{i} d \protect\mu_{j}$}

\label{sec:dmudmu}

We first calculate the following,
\begin{eqnarray}
&&h^{2}r^{2}d\mu _{3}^{2}=\frac{1}{\sqrt{\Delta }}\frac{r^{2}\Delta }{%
r^{2}\Delta +\mu _{3}^{2}}d\mu _{3}^{2}  \notag \\
&=&\frac{1}{\sqrt{\Delta }}d\mu _{3}^{2}-\frac{1}{\sqrt{\Delta }}\frac{1}{%
r^{2}\Delta +\mu _{3}^{2}}(\mu _{1}d\mu _{1}+\mu _{2}d\mu _{2})^{2}  \notag
\\
&=&\frac{1}{\sqrt{\Delta }}d\mu _{3}^{2}-\frac{h^{2}}{\Delta }(\mu
_{1}^{2}d\mu _{1}^{2}+\mu _{2}^{2}d\mu _{2}^{2}+2\mu _{1}\mu _{2}d\mu
_{1}d\mu _{2}).  \label{dmu1}
\end{eqnarray}%
Using this, we have
\begin{eqnarray}
&&h^{2}\left( r^{2}d\mu _{3}^{2}+S_{11}R_{1}^{2}d\mu
_{1}^{2}+S_{22}R_{2}^{2}d\mu _{2}^{2}+2S_{12}R_{1}R_{2}d\mu _{1}d\mu
_{2}\right)  \notag \\
&=&\frac{1}{\sqrt{\Delta }}\left( d\mu _{3}^{2}+H_{1}d\mu _{1}^{2}+H_{2}d\mu
_{2}^{2}\right) +2h^{2}\left( S_{12}R_{1}R_{2}-\frac{\mu _{1}\mu _{2}}{%
\Delta }\right) d\mu _{1}d\mu _{2},  \notag \\
&&  \label{muterms}
\end{eqnarray}%
where we have defined
\begin{equation}
H_{i}=\sqrt{\Delta }h^{2}\left( S_{ii}R_{i}^{2}-\frac{\mu _{i}^{2}}{\Delta }%
\right)  \label{defofH}
\end{equation}%
for $i=1,2$. For two-charge superstar, this becomes the $H_{i}$ of the
superstar.

When the following condition is satisfied,
\begin{equation}
S_{12}R_{1}R_{2}-\frac{\mu _{1}\mu _{2}}{\Delta }=0,  \label{nodmu1dmu2}
\end{equation}%
the metric does not have the mixing term $d\mu _{1}d\mu _{2}$. For
superstar, this is the case.


\subsection{Terms with $d \protect\phi_{i} d \protect\phi_{j}$}

\label{sec:dphidphi}

The relevant terms in the metric (\ref{NewCoordinateMetricBeforeShift}) are
\begin{eqnarray}
&&\sum_{i=1,2}(-h^{-2}2\omega _{\phi _{i}}dtd\phi _{i}-h^{-2}\omega _{\phi
_{i}}^{2}d\phi _{i}^{2}+h^{2}S_{ii}r_{i}^{2}d\phi _{i}^{2})  \notag \\
&&\quad -h^{-2}2\omega _{\phi _{1}}\omega _{\phi _{2}}d\phi _{1}d\phi
_{2}+2h^{2}S_{12}R_{1}R_{2}\mu _{1}\mu _{2}d\phi _{1}d\phi _{2}  \notag \\
&=&h^{-2}\left( S_{1}d\phi _{1}^{2}-2\omega _{\phi _{1}}dtd\phi
_{1}+S_{2}d\phi _{2}^{2}-2\omega _{\phi _{2}}dtd\phi _{2}+2N_{12}d\phi
_{1}d\phi _{2}\right) ,  \label{dphidtterms}
\end{eqnarray}%
where we have defined
\begin{eqnarray}
&&S_{i}=h^{4}S_{ii}r_{i}^{2}-\omega _{\phi _{i}}^{2}\text{~}~~(i=1,2), \\
&&N_{12}=h^{4}S_{12}r_{1}r_{2}-\omega _{\phi _{1}}\omega _{\phi _{2}}.
\end{eqnarray}

Making the shift $\phi _{i}\rightarrow \phi _{i}-t$, (\ref{dphidtterms})
becomes
\begin{eqnarray}
&&h^{-2}S_{1}d\phi _{1}^{2}-2h^{-2}(S_{1}+\omega _{\phi _{1}}+N_{12})dtd\phi
_{1}  \notag \\
&&+h^{-2}S_{2}d\phi _{2}^{2}-2h^{-2}(S_{2}+\omega _{\phi
_{2}}+N_{12})dtd\phi _{2}+2h^{-2}N_{12}d\phi _{1}d\phi _{2}  \notag \\
&&+h^{-2}(S_{1}+S_{2}+2\omega _{\phi _{1}}+2\omega _{\phi
_{2}}+2N_{12})dt^{2}  \notag \\
&=&h^{-2}\left( S_{1}d\phi _{1}^{2}+2S_{1t}dtd\phi _{2}+S_{2}d\phi
_{2}^{2}+2S_{2t}dtd\phi _{3}+2N_{12}d\phi _{1}d\phi _{2}+S_{t}dt^{2}\right) .
\label{shiftdphidtterms}
\end{eqnarray}%
where we have defined
\begin{eqnarray}
&&S_{1t}=-S_{1}-\omega _{\phi _{1}}-N_{12},  \notag \\
&&S_{2t}=-S_{2}-\omega _{\phi _{2}}-N_{12},  \notag \\
&&S_{t}=S_{1}+S_{2}+2\omega _{\phi _{1}}+2\omega _{\phi _{2}}+2N_{12}.
\end{eqnarray}

Finally, (\ref{shiftdphidtterms}) can be written as
\begin{eqnarray}
&& S_{1}d\phi_{1}^{2}+2S_{1t}dtd\phi_{1}
+S_{2}d\phi_{2}^{2}+2S_{2t}dtd\phi_{2} +2N_{12}d\phi_{1}d\phi_{2}
+S_{t}dt^{2}  \notag \\
&=& h_{ij}(d \phi_{i}+M_{i}dt)(d \phi_{j}+M_{j}dt)-h_{ij}M_{i}M_{j}dt^{2}+
S_{t}dt^{2},
\end{eqnarray}
where
\begin{eqnarray}
&&h_{11}=S_{1},\quad h_{22}=S_{2},\quad h_{12}=N_{12} ,
\end{eqnarray}
and
\begin{eqnarray}
\left(
\begin{array}{c}
M_{1} \\
M_{2}%
\end{array}%
\right) &=&\frac{1}{S_{1}S_{2}-N_{12}^{2}}\left(
\begin{array}{c}
S_{2}S_{1t}-N_{12}S_{2t} \\
S_{1}S_{2t}-N_{12}S_{1t}%
\end{array}%
\right)  \notag \\
&=& -1+ \frac{1}{S_{1}S_{2}-N_{12}^{2}}\left(
\begin{array}{c}
-S_{2}\omega_{\phi_{1}}+N_{12}\omega_{\phi_{2}} \\
-S_{1}\omega_{\phi_{2}}+N_{12}\omega_{\phi_{1}}%
\end{array}%
\right).  \label{M_i}
\end{eqnarray}

When $h_{12}=N_{12}=0$, the metric does not have the mixing term $d\phi
_{1}d\phi _{2}$, and $M_{i}$ will get a simple expression. Defining $%
N_{i}=M_{i}|_{N_{12}=0}$, we have
\begin{eqnarray}
N_{i} &=&-1-S_{i}^{-1}\omega _{\phi _{i}}  \notag \\
&=&-1-\frac{\omega _{\phi _{i}}}{h^{4}S_{ii}r_{i}^{2}-\omega _{\phi _{i}}^{2}%
}.  \label{shiftin1/4}
\end{eqnarray}%
For superstar, $N_{12}=0$, and $N_{i}$ are calculated
\begin{equation}
N_{i}=-\frac{q_{i}}{r^{2}+q_{i}}.
\end{equation}%
See Appendix \ref{formulaads} for more details.



\subsection{Terms with $d \protect\mu _{i} d r$}

\label{sec:analysisondmudr}

Here we will analyze the condition under which the mixing terms $d\mu _{i}dr$
vanish.

The mixing terms vanish when the following conditions are satisfied
\begin{eqnarray}
&&S_{11}R_{1}T_{1}\mu _{1}+S_{12}T_{2}\mu _{2}R_{1}=r\mu _{1},  \notag \\
&&S_{22}R_{2}T_{2}\mu _{2}+S_{12}T_{1}\mu _{1}R_{2}=r\mu _{2},
\label{demandforzerodmudr}
\end{eqnarray}%
One can show that these are satisfied for two-charge superstar with the help
of the equations in Appendix \ref{formulaads}.

The set of the equations can be summarized as
\begin{equation}
\left(
\begin{array}{cc}
S_{11}R_{1}\mu _{1} & S_{12}R_{1}\mu _{2} \\
S_{12}R_{2}\mu _{1} & S_{22}R_{2}\mu _{2}%
\end{array}%
\right) \left(
\begin{array}{c}
T_{1} \\
T_{2}%
\end{array}%
\right) =r\left(
\begin{array}{c}
\mu _{1} \\
\mu _{2}%
\end{array}%
\right)
\end{equation}%
Note that the determinant of the matrix is $%
(S_{11}S_{22}-S_{12}^{2})R_{1}R_{2}\mu _{1}\mu _{2}=(\det S_{ij}) r_{1}r
_{2} $. Using $T_{i}=R_{i}^{\prime }$,
\begin{eqnarray}
\left(
\begin{array}{c}
R_{1}^{\prime } \\
R_{2}^{\prime }%
\end{array}%
\right) &=&\frac{r}{(\det S_{ij})r _{1}r _{2}}\left(
\begin{array}{cc}
S_{22}R_{2}\mu _{2} & -S_{12}R_{1}\mu _{2} \\
-S_{12}R_{2}\mu _{1} & S_{11}R_{1}\mu _{1}%
\end{array}%
\right) \left(
\begin{array}{c}
\mu _{1} \\
\mu _{2}%
\end{array}%
\right)  \notag \\
&=&\frac{r}{(\det S_{ij})r_{1}r_{2}}\left(
\begin{array}{c}
\mu _{2}(S_{22}R_{2}\mu _{1}-S_{12}R_{1}\mu _{2}) \\
\mu _{1}(S_{11}R_{1}\mu _{2}-S_{12}R_{2}\mu _{1})%
\end{array}%
\right)  \notag \\
&=&\frac{r}{(\det S_{ij})}\left(
\begin{array}{c}
S_{22}R_{1}^{-1}-R_{12}R_{2}^{-1}\mu _{2}\mu _{1}^{-1} \\
S_{11}R_{2}^{-1}-R_{12}R_{1}^{-1}\mu _{1}\mu _{2}^{-1}%
\end{array}%
\right)  \label{renritudiffequation}
\end{eqnarray}

It can be further simplified if we use (\ref{nodmu1dmu2}), which is the
equation for vanishing $d\mu _{1}d\mu _{2}~$term. We then have
\begin{eqnarray}
\left(
\begin{array}{c}
R_{1}^{\prime } \\
R_{2}^{\prime }%
\end{array}%
\right) &=&\frac{r}{(\det S_{ij})}\left(
\begin{array}{c}
S_{22}R_{1}^{-1}-\frac{1}{\Delta }R_{1}^{-1}R_{2}^{-2}\mu _{2}^{2} \\
S_{11}R_{2}^{-1}-\frac{1}{\Delta }R_{2}^{-1}R_{1}^{-2}\mu _{1}^{2}%
\end{array}%
\right)  \notag \\
&=&\frac{r}{(\det S_{ij})}\left(
\begin{array}{c}
\frac{1}{\sqrt{\Delta }}h^{-2}H_{2}R_{1}^{-1}R_{2}^{-2} \\
\frac{1}{\sqrt{\Delta }}h^{-2}H_{1}R_{2}^{-1}R_{1}^{-2}%
\end{array}%
\right) .
\end{eqnarray}%
where we have used the definition of $H_{i}$ (\ref{defofH}). Dividing the
first equation by the second equation, we obtain
\begin{equation}
(\log R_{1}^{2})^{\prime }H_{1}=(\log R_{2}^{2})^{\prime }H_{2}
\label{eqn_2}
\end{equation}%
We have that for superstar, (\ref{eqn_2}) is equal to $2rH_{1}H_{2}f^{-1}$.



\vspace{1pt}

\section{Derivation of the $N_{1 2}$, $M_ 1$, $M_ 2$ in large $r$}

\vspace{1pt}

\vspace{1pt}\renewcommand{\theequation}{C.\arabic{equation}} %
\setcounter{equation}{0}

\vspace{1pt}

\label{appendix_N12_M1_M2}

\vspace{1pt}

\vspace{1pt}

\vspace{1pt}

\vspace{1pt}

In this appendix, we present the details of the calculations to derive the
asymptotic forms of the mixing functions $N_{12}$, $M_{1}$, $M_{2}$.

Using (\ref{K_alpha_0}),(\ref{K_kappa_1}),(\ref{K_kappa_2}),(\ref{K_alpha_2}%
), the expressions (\ref{def-of-sn}) are evaluated as
\begin{eqnarray}
n_{12} &=&2K_{2}\partial _{t_{1}}\partial _{t_{2}}K_{0}-\partial
_{t_{1}}K_{1}\partial _{t_{2}}K_{1}  \notag \\
&=&\frac{q}{2a^{2}R^{4}}\left( 1+\frac{q}{aR^{2}}(2+\alpha
_{2})+O(1/r^{4})\right) \frac{2qr_{1}^{2}r_{2}^{2}}{R^{4}}\left( 1+\frac{%
q\alpha _{0}}{aR^{2}}+O(1/r^{4})\right)  \notag \\
&&-r_{1}^{2}r_{2}^{2}\left( \frac{q}{aR^{4}}\right) ^{2}\left( 1+\frac{q}{%
aR^{2}}(1+\kappa _{1})+O(1/r^{4})\right) \left( 1+\frac{q}{aR^{2}}(1+\kappa
_{2})+O(1/r^{4})\right)  \notag \\
&=&\frac{r_{1}^{2}r_{2}^{2}}{a^{2}R^{8}}\left( \frac{q^{3}}{aR^{2}}\left(
\alpha _{2}+\alpha _{0}-\kappa _{1}-\kappa _{2}\right) +O(1/r^{4})\right)
\notag \\
&=&\frac{q^{3}r_{1}^{2}r_{2}^{2}}{a^{3}R^{10}}k+O(1/r^{8}),
\end{eqnarray}%
and
\begin{eqnarray}
s_{1} &=&2K_{2}\partial _{t_{1}}^{2}K_{0}-(\partial _{t_{1}}K_{1})^{2}
\notag \\
&=&\frac{q}{2a^{2}R^{4}}\left( 1+\frac{q(\alpha _{2}+2)}{aR^{2}}\right)
\left( 2ar_{1}^{2}-\frac{2qr_{1}^{2}r_{2}^{2}}{R^{4}}+O(1/r^{2})\right)
-\left( \frac{qr_{1}^{2}}{aR^{4}}\right) ^{2}\left( 1+\frac{q}{aR^{2}}%
(1+\kappa _{1})\right) ^{2}  \notag \\
&=&\frac{qr_{1}^{2}}{aR^{4}}\left( 1+\frac{q(\alpha _{2}+2)}{aR^{2}}-\frac{%
qr_{2}^{2}}{aR^{4}}-\frac{qr_{1}^{2}}{aR^{4}}+O(1/r^{4})\right)  \notag \\
&=&\frac{qr_{1}^{2}}{aR^{4}}\left( 1+\frac{q(\alpha _{2}+1)}{aR^{2}}%
+O(1/r^{4})\right) ,  \notag \\
s_{2} &=&\frac{qr_{2}^{2}}{aR^{4}}\left( 1+\frac{q(\alpha _{2}+1)}{aR^{2}}%
+O(1/r^{4})\right) .
\end{eqnarray}

The mixing between time and the angles are given by $M_{i}$
\begin{equation}
M_{1}=-1+\frac{s_{2}\partial _{t_{1}}K_{1}-n_{12}\partial _{t_{2}}K_{1}}{%
s_{1}s_{2}-n_{12}^{2}},
\end{equation}%
\begin{equation}
M_{2}=-1+\frac{s_{1}\partial _{t_{2}}K_{1}-n_{12}\partial _{t_{1}}K_{1}}{%
s_{1}s_{2}-n_{12}^{2}}.
\end{equation}%
Since $s_{1},s_{2},\partial _{t_{1}}K_{1},\partial _{t_{2}}K_{1}=O(1/r^{2}),$
and $n_{12}=O(k/r^{6})$, the effect of $n_{12}^{2}$ in the denominator is to
give at most $O(1/r^{8})$ terms in $M_{1},M_{2}.$ So the following
expressions are up to $O(1/r^{8})$,%
\begin{equation}
M_{1}=-1+\frac{\partial _{t_{1}}K_{1}}{s_{1}}-\frac{n_{12}\partial
_{t_{2}}K_{1}}{s_{1}s_{2}}+O(1/r^{8}),
\end{equation}%
\begin{equation}
M_{2}=-1+\frac{\partial _{t_{2}}K_{1}}{s_{2}}-\frac{n_{12}\partial
_{t_{1}}K_{1}}{s_{1}s_{2}}+O(1/r^{8}),
\end{equation}%
where the last terms will be evaluated as
\begin{equation}
-\frac{n_{12}\partial _{t_{2}}K_{1}}{s_{1}s_{2}}=-\frac{q^{2}r_{2}^{2}}{%
a^{2}R^{6}}k+O(1/r^{6}),
\end{equation}%
\begin{equation}
-\frac{n_{12}\partial _{t_{1}}K_{1}}{s_{1}s_{2}}=-\frac{q^{2}r_{1}^{2}}{%
a^{2}R^{6}}k+O(1/r^{6}).
\end{equation}

In other words, the difference between $M_{i},N_{i}$ is
\begin{equation}
M_{1}-N_{1}=-\frac{n_{12}\partial _{t_{2}}K_{1}}{s_{1}s_{2}}%
+O(1/r^{8})=\left( -k\frac{q^{2}r_{2}^{2}}{a^{2}R^{6}}+O(1/r^{6})\right)
+O(1/r^{8}),
\end{equation}%
\begin{equation}
M_{2}-N_{2}=-\frac{n_{12}\partial _{t_{1}}K_{1}}{s_{1}s_{2}}%
+O(1/r^{8})=\left( -k\frac{q^{2}r_{1}^{2}}{a^{2}R^{6}}+O(1/r^{6})\right)
+O(1/r^{8}).
\end{equation}%
The $-k\frac{q^{2}r_{2}^{2}}{a^{2}R^{6}}$ is one of the $O(1/r^{4})$ terms
in $N_{1}$ and in $M_{1}$, but it is the only $O(1/r^{4})$ term in $%
M_{1}-N_{1}$. Therefore$~M_{i}-N_{i}=O(k/r^{4}),$ for $k\neq 0$; and $%
M_{i}-N_{i}=O(1/r^{6}),$ for$~k=0.$

Now we calculate $M_{i}$ in the large $r$,
\begin{eqnarray}
M_{1} &=&-1+\frac{\partial _{t_{1}}K_{1}}{s_{1}}+O(1/r^{4})  \notag \\
&=&-1+\left( 1+\frac{q}{aR^{2}}(1+\kappa _{1})\right) \left( 1+\frac{%
q(\alpha _{2}+1)}{aR^{2}}+O(1/r^{4})\right) ^{-1}+O(1/r^{4})  \notag \\
&=&\frac{q}{aR^{2}}\left( -\alpha _{2}+\kappa _{1}\right) +O(1/r^{4})  \notag
\\
&=&\frac{q}{r^{2}}\left( -\alpha _{2}+\kappa _{1}\right) +O(1/r^{4}).
\end{eqnarray}%
Similarly
\begin{eqnarray}
M_{2} &=&-1+\frac{\partial _{t_{2}}K_{1}}{s_{2}}+O(1/r^{4})  \notag \\
&=&-1+\left( 1+\frac{q}{aR^{2}}(1+\kappa _{2})\right) \left( 1+\frac{%
q(\alpha _{2}+1)}{aR^{2}}+O(1/r^{4})\right) ^{-1}+O(1/r^{4})  \notag \\
&=&\frac{q}{aR^{2}}\left( -\alpha _{2}+\kappa _{2}\right) +O(1/r^{4})  \notag
\\
&=&\frac{q}{r^{2}}\left( -\alpha _{2}+\kappa _{2}\right) +O(1/r^{4}).
\end{eqnarray}

Now we focus on the family of solutions\vspace{1pt}

\begin{equation}
K_{0}^{(1)}=\frac{q^{2}}{a}\left( \frac{dr_{1}^{2}+er_{2}^{2}}{R^{4}}+\frac{%
fr_{1}r_{2}}{R^{4}}\right) .
\end{equation}

\vspace{1pt}1. when $f=0$,
\begin{eqnarray}
&&\kappa _{1}=\frac{4(e-d)}{R^{2}}(2r_{2}^{2}-r_{1}^{2}),  \notag \\
&&\kappa _{2}=\frac{4(d-e)}{R^{2}}(2r_{1}^{2}-r_{2}^{2}),  \notag \\
&&\alpha _{0}=\frac{4}{R^{2}}\left(
d(2r_{1}^{2}-r_{2}^{2})+e(2r_{2}^{2}-r_{1}^{2})\right) .
\end{eqnarray}%
Note that $\kappa _{1}=\kappa _{2}=0$, $\alpha _{0}\neq 0$ when $d=e$.

2. when $d=e=0$,
\begin{eqnarray}
\kappa _{1} &=&-\frac{f}{r_{1}r_{2}R^{4}}%
(r_{2}^{6}+13r_{1}^{2}r_{2}^{4}-33r_{1}^{4}r_{2}^{2}+3r_{1}^{6}),  \notag \\
\kappa _{2} &=&-\frac{f}{r_{1}r_{2}R^{4}}%
(r_{1}^{6}+13r_{2}^{2}r_{1}^{4}-33r_{2}^{4}r_{1}^{2}+3r_{2}^{6}),  \notag \\
\alpha _{0} &=&-\frac{3f}{r_{1}r_{3}R^{2}}%
(r_{1}^{4}-6r_{2}^{2}r_{1}^{4}+r_{2}^{4}).
\end{eqnarray}%
Although it is possible to have the third parameter $f,$ we think
that it does not have a relevant physical meaning, and we have
chosen $f=0$. We have set therefore
\begin{equation}
f=0.
\end{equation}

These are the solutions presented in the text in (\ref{K_0_(1)_d_e})-(\ref%
{K_(1)_d_e}),(\ref{solution_d_e_02}).\vspace{1pt}

From the expressions of $N_{12},M_{1},M_{2},$\
\begin{eqnarray}
q\left( \alpha _{2}-\kappa _{1}\right) &=&qm=q_{1},  \notag \\
q\left( \alpha _{2}-\kappa _{2}\right) &=&qn=q_{2},  \notag \\
\alpha _{0}+\alpha _{2}-\kappa _{1}-\kappa _{2} &=&k.
\end{eqnarray}%
Plugging these conditions into the solutions (\ref{solution_d_e_02}), we find%
\begin{eqnarray}
4e &=&k-m=k-q_{1}/q,  \notag \\
4d &=&k-n=k-q_{2}/q,  \notag \\
c &=&k.
\end{eqnarray}%
and we have $\kappa _{1}-\kappa _{2}=n-m.$

There are several equivalent and alternative ways of writing these
variables. For example,
\begin{eqnarray}
q\alpha _{0} &=&q\alpha _{2}+qk-q_{1}-q_{2},  \notag \\
q\alpha _{2} &=&q\alpha _{0}-qk+q_{1}+q_{2},  \notag \\
q\kappa _{1} &=&q\alpha _{2}-q_{1}=q\alpha _{0}-qk+q_{2},  \notag \\
q\kappa _{2} &=&q\alpha _{2}-q_{2}=q\alpha _{0}-qk+q_{1},  \notag \\
qk &=&q\alpha _{0}-q\alpha _{2}+q_{1}+q_{2}=q\alpha _{0}+q\alpha
_{2}-q\kappa _{1}-q\kappa _{2},  \notag \\
2q\alpha _{2} &=&q\kappa _{1}+q\kappa _{2}+q_{1}+q_{2},  \notag \\
q_{1}+q_{2} &=&q\alpha _{2}-q\alpha _{0}+qk=2q\alpha _{2}-q\kappa
_{1}-q\kappa _{2},  \notag \\
q\kappa _{1}-q\kappa _{2} &=&q_{2}-q_{1},  \notag \\
q\kappa _{1}+q\kappa _{2} &=&q\alpha _{0}+q\alpha _{2}-qk=2q\alpha
_{2}-q_{1}-q_{2}.
\end{eqnarray}

\vspace{1pt}


\vspace{1pt}

\vspace{1pt}

\vspace{1pt}

\vspace{1pt}

\section{Derivation of asymptotic metric}

\vspace{1pt}\renewcommand{\theequation}{D.\arabic{equation}} %
\setcounter{equation}{0}

\vspace{1pt}

\label{appendix_derivation_metric_r_large}

\vspace{1pt}

In this appendix, we present the details of the derivation of (\ref%
{asymptoticmetric1}) and (\ref{asymptoticmetric2}).


\subsection{Metric functions}

\vspace{1pt} Here we collect equations which would be helpful to calculate
the asymptotic form of the metric.

\begin{equation}
h^{2}=\frac{1}{r^{2}\sqrt{\Delta }}+O(y^{2}),  \label{h2expand}
\end{equation}%
\begin{equation}
\Delta =\frac{1}{4r^{4}K_{2}}+O(y^{2}).  \label{DeltabyK2}
\end{equation}%
\begin{eqnarray}
K_{2} &=&\frac{q}{4a^{2}R^{4}}\left( 1+\frac{q(\alpha _{2}+2)}{aR^{2}}%
\right) +O(1/r^{8})  \notag \\
&=&\frac{q}{4r^{4}}\left( 1-2\frac{qC_{1}\mu _{1}^{2}+qC_{2}\mu _{2}^{2}}{%
r^{2}}\right) \left( 1+\frac{q(\alpha _{2}+2)}{r^{2}}\right) +O(1/r^{8})
\notag \\
&=&\frac{q}{4r^{4}}\left( 1+\frac{q(\alpha _{2}+2-2qC_{1}\mu
_{1}^{2}-2qC_{2}\mu _{2}^{2})}{r^{2}}\right) +O(1/r^{8}).
\label{K2expansion}
\end{eqnarray}

\begin{equation}
R_{1}=\frac{1}{\sqrt{a}}\sqrt{r^{2}+qC_{1}}\left( 1+O(1/r^{2})\right)
,~\quad R_{2}=\frac{1}{\sqrt{a}}\sqrt{r^{2}+qC_{2}}\left(
1+O(1/r^{2})\right) ,
\end{equation}%
\begin{equation}
aR^{2}=a(r_{1}^{2}+r_{2}^{2})=r^{2}+qC_{1}\mu _{1}^{2}+qC_{2}\mu
_{2}^{2}+O(1/r^{2})-O(y^{2}).
\end{equation}

\begin{eqnarray}
S_{ii} &=&2\frac{\partial _{i}\partial _{\bar{\imath}}K}{Z+\frac{1}{2}}=%
\frac{1}{2}\frac{1}{r_{i}^{2}}\partial _{t_{i}}^{2}K+O(y^{2})  \notag \\
&=&a-\frac{q(R^{2}-r_{i}^{2})}{R^{4}}+O(1/r^{4})+O(y^{2}),
\end{eqnarray}%
\begin{eqnarray}
S_{12} &=&\frac{1}{2}\frac{\partial _{r_{1}}\partial _{r_{2}}K}{Z+\frac{1}{2}%
}=\frac{1}{2}\partial _{r_{1}}\partial _{r_{2}}K_{0}+O(y^{2})  \notag \\
&=&\frac{qr_{1}r_{2}}{R^{4}}\left( 1+\frac{q\alpha _{0}}{aR^{2}}%
+O(1/r^{4})\right) +O(y^{2}).
\end{eqnarray}%
\begin{equation*}
s_{i}=\frac{qr_{i}^{2}}{aR^{4}}\left( 1+\frac{q(\alpha _{2}+1)}{aR^{2}}%
+O(1/r^{4})\right) ,
\end{equation*}%
\begin{equation}
h^{2}=\sqrt{4K_{2}}+O(y^{2})=\frac{\sqrt{q}}{aR^{2}}\left( 1+\frac{q(\alpha
_{2}+2)}{2aR^{2}}\right) +O(1/r^{6})+O(y^{2}).
\end{equation}

Comparing with (\ref{expandZat1/2}),(\ref{Z_exp_delta}),
\begin{equation}
K_{2}=\frac{r_{0}^{2}}{4r^{4}\Delta }.
\end{equation}%
Because AdS is given by $\Delta =1$, $K_{2}{}_{(AdS)}=\frac{r_{0}^{2}}{4r^{4}%
}$.$~$We also have,$~$%
\begin{equation}
K_{2}=\frac{q}{4(aR^{2}-q)^{2}},
\end{equation}%
for AdS. So we have that for AdS, $C_{1}=C_{2}=1$. Note that $%
R^{2}=r_{1}^{2}+r_{2}^{2}$.

%
%
%
%
%
%
%
%
%
%
%
%
%
%
%
\vspace{1pt}



\subsection{Calculation of metric}

The factor in front of $dt^{2}$ is now calculated. Taking account of $%
M_{i}=O\left( 1/r^{2}\right) $ and $N_{12}=O\left( 1/r^{6}\right) $, we have
\begin{eqnarray}
&&h^{-2}\left( 1+h_{ab}M_{a}M_{b}-S_{t}\right)  \notag
\label{calculationdt^2} \\
&=&h^{-2}\left(
1+S_{1}M_{1}^{2}+S_{2}M_{2}^{2}+2N_{12}M_{1}M_{2}-S_{1}-S_{2}-2\omega
_{1}-2\omega _{2}-2N_{12}\right)  \notag \\
&=&h^{-2}\left( 1-S_{1}-S_{2}-2\omega _{1}-2\omega _{2}+O(1/r^{6})\right)
\notag \\
&=&h^{-2}\left( 1-\frac{qr_{1}^{2}}{aR^{4}}\left( 1+\frac{q(\alpha _{2}+1)}{%
aR^{2}}\right) -\frac{qr_{2}^{2}}{aR^{4}}\left( 1+\frac{q(\alpha _{2}+1)}{%
aR^{2}}\right) \right.  \notag \\
&&\left. +2\frac{qr_{1}^{2}}{aR^{4}}\left( 1+\frac{q(1+\kappa _{1})}{aR^{2}}%
\right) +2\frac{qr_{2}^{2}}{aR^{4}}\left( 1+\frac{q(1+\kappa _{2})}{aR^{2}}%
\right) +O(1/r^{6})+O(y^{2})\right)  \notag \\
&=&\sqrt{\Delta }r^{2}\left( 1+\frac{q}{aR^{2}}+\frac{q^{2}(1-\alpha _{2})}{%
a^{2}R^{4}}+\frac{2q^{2}(r_{1}^{2}\kappa _{1}+r_{2}^{2}\kappa _{2})}{%
a^{2}R^{6}}\right) +O(1/r^{4})+O(y^{2})  \notag \\
&=&\sqrt{\Delta }\left( r^{2}+q-q^{2}\frac{C_{1}\mu _{1}^{2}+C_{2}\mu
_{2}^{2}}{r^{2}}+\frac{q^{2}(1-\alpha _{2})}{r^{2}}+\frac{2q^{2}(\kappa
_{1}\mu _{1}^{2}+\kappa _{2}\mu _{2}^{2})}{r^{2}}\right)
+O(1/r^{4})+O(y^{2}).  \notag \\
&&
\end{eqnarray}

We next calculate the factor in front of $dr^2$ in large $r$. We first show
\begin{eqnarray}
&&\mu _{3}^{2}+S_{11}\mu _{1}^{2}T_{1}^{2}
+S_{2}\mu_{2}^{2}T_{2}^{2}+2S_{12}\mu _{1}\mu _{2}T_{1}T_{2}  \notag \\
&=&\mu _{3}^{2}+a \left( 1-\frac{qr _{2}^{2}}{aR^{4}}\right) \mu
_{1}^{2}T_{1}^{2} +a\left( 1-\frac{qr_{1}^{2}}{aR^{4}}\right) \mu
_{2}^{2}T_{2}^{2} +2\frac{qr _{1}r _{2}}{R^{4}}\mu _{1}\mu _{2}T_{1}T_{2}
+O(1/r^4)  \notag \\
&=&\mu _{3}^{2}+a\mu _{1}^{2}T_{1}^{2}+a\mu _{2}^{2}T_{2}^{2}-\frac{q}{R^{4}}%
\left( r_{1}\mu _{2}T_{2}-r _{2}\mu _{1}T_{1}\right) ^{2} +O(1/r^4)  \notag
\\
&= &\mu _{3}^{2}+\mu _{1}^{2}\left( 1-\frac{qC_{1}}{r^{2}}\right) +\mu
_{2}^{2}\left( 1-\frac{qC_{2}}{r^{2}}\right) +O(1/r^4)  \notag \\
&=& 1-\frac{qC_{1}\mu _{1}^{2}+qC_{2}\mu _{2}^{2}}{r^{2}} +O(1/r^4).
\end{eqnarray}%
With the help of (\ref{h2expand}), (\ref{DeltabyK2}) and (\ref{K2expansion}%
), we obtain
\begin{eqnarray}
&&h^{2}\left( \mu _{3}^{2}+S_{11}\mu _{1}^{2}T_{1}^{2}+S_{22}\mu
_{2}^{2}T_{2}^{2}+2S_{12}\mu _{1}\mu _{2}T_{1}T_{2}\right)  \notag \\
&=&\sqrt{\Delta }4r^{2}K_{2}\left( 1-\frac{qC_{1}\mu _{1}^{2}+qC_{2}\mu
_{2}^{2}}{r^{2}}\right) +O(1/r^4)+O(y^2)  \notag \\
&=&\sqrt{\Delta }\frac{q}{r^{2}}\left( 1+\frac{q(\alpha_{2}+2)-3q(C_{1}\mu
_{1}^{2}+C_{2}\mu _{2}^{2})}{r^{2}} \right) +O(1/r^4) +O(y^2).
\end{eqnarray}

The factor $H_{i}$ is evaluated at large $r$,
\begin{eqnarray}
H_{1} &=&\sqrt{\Delta }h^{2}\left( S_{11}R_{1}^{2}-\frac{\mu _{1}^{2}}{%
\Delta }\right)  \notag \\
&=&\frac{1}{r^{2}}\left( \left( a-\frac{qr_{2}^{2}}{R^{4}}\right)
R_{1}^{2}-4r^{4}K_{2}\mu _{1}^{2}\right) +O(1/r^{4})+O(y^{2})  \notag \\
&=&\frac{1}{r^{2}}\left( \left( a-\frac{qr_{2}^{2}}{R^{4}}\right)
R_{1}^{2}-r^{4}\frac{q}{a^{2}R^{4}}\mu _{1}^{2}\right) +O(1/r^{4})+O(y^{2})
\notag \\
&=&\frac{1}{r^{2}}\left( \left( 1-\frac{q\mu _{2}^{2}}{r^{2}}\right)
(r^{2}+qC_{1})-q\mu _{1}^{2}\right) +O(1/r^{4})+O(y^{2})  \notag \\
&=&1+\frac{q(C_{1}-1)}{r^{2}}+O(1/r^{4})+O(y^{2}), \\
H_{2} &=&1+\frac{q(C_{2}-1)}{r^{2}}+O(1/r^{4})+O(y^{2}).
\end{eqnarray}

We calculate the factor in front of $d\mu _{1}d\mu _{2}$ in large $r$:
\begin{eqnarray}
&&2h^{2}\left( S_{12}R_{1}R_{2}-\frac{\mu _{1}\mu _{2}}{\Delta }\right)
\notag \\
&=&\frac{2}{r^{2}\sqrt{\Delta }}\left( \frac{qr_{1}r_{2}}{R^{4}}\left( 1+%
\frac{q\alpha _{0}}{aR^{2}}+O(1/r^{4})\right) R_{1}R_{2}-\mu _{1}\mu
_{2}4r^{4}K_{2}\right) +O(y^{2})  \notag \\
&=&\frac{2}{r^{2}\sqrt{\Delta }}\left( \frac{q}{R^{4}}\left( 1+\frac{q\alpha
_{0}}{aR^{2}}+O(1/r^{4})\right) R_{1}^{2}R_{2}^{2}-4r^{4}K_{2}\right) \mu
_{1}\mu _{2}+O(y^{2})  \notag \\
&=&\frac{q}{a^{2}R^{4}}\frac{2}{r^{2}\sqrt{\Delta }}\left( a^{2}\left( 1+%
\frac{q\alpha _{0}}{aR^{2}}\right) R_{1}^{2}R_{2}^{2}-r^{4}\left( 1+\frac{%
q(\alpha _{2}+2)}{r^{2}}\right) +O(1)\right) \mu _{1}\mu _{2}+O(y^{2})
\notag \\
&=&\frac{2}{\sqrt{\Delta }}\frac{q^{2}}{r^{4}}\left( C_{1}+C_{2}+\alpha
_{0}-\alpha _{2}-2\right) \mu _{1}\mu _{2}+O(1/r^{6})+O(y^{2}).
\end{eqnarray}

In order to calculate the factor in front of $d\mu _{i}dr$, we calculate
\begin{eqnarray}
&&S_{11}R_{1}T_{1}\mu _{1}+S_{12}T_{2}\mu _{2}R_{1}  \notag \\
&=&\left( a-\frac{qr_{2}^{2}}{R^{4}}+O(1/r^{4})\right) R_{1}T_{1}\mu
_{1}+\left( \frac{qr_{1}r_{2}}{R^{4}}+O(1/r^{4})\right) T_{2}\mu
_{2}R_{1}+O(y^{2})  \notag \\
&=&aR_{1}T_{1}\mu _{1}-\frac{qr_{2}^{2}}{R^{4}}R_{1}T_{1}\mu _{1}+\frac{%
qr_{1}r_{2}}{R^{4}}T_{2}\mu _{2}R_{1}+O(1/r^{3})+O(y^{2})  \notag \\
&=&aR_{1}T_{1}\mu _{1}+\frac{q\mu _{1}\mu _{2}^{2}}{R^{4}}%
(-R_{1}T_{1}R_{2}^{2}+R_{2}T_{2}R_{1}^{2})+O(1/r^{3})+O(y^{2})  \notag \\
&=&r\mu _{1}+\frac{q\mu _{1}\mu _{2}^{2}}{R^{4}}\left( -\frac{r}{a}R_{2}^{2}+%
\frac{r}{a}R_{1}^{2}\right) +O(1/r^{3})+O(y^{2})  \notag \\
&=&r\mu _{1}+\frac{q^{2}\mu _{1}\mu _{2}^{2}}{r^{3}}\left(
C_{1}-C_{2}\right) +O(1/r^{3})+O(y^{2}).
\end{eqnarray}%
Note that $R_{1}T_{1}=r/a+O(1/r^{3})$. Therefore, we have
\begin{eqnarray}
&&h^{2}(S_{11}R_{1}T_{1}\mu _{1}+S_{12}T_{2}\mu _{2}R_{1}-r\mu _{1})  \notag
\\
&=&\frac{1}{r^{2}\sqrt{\Delta }}O(1/r^{3})+O(y^{2})  \notag \\
&=&\frac{1}{\sqrt{\Delta }}O(1/r^{5})+O(y^{2}).
\end{eqnarray}

The functions in front of $d\phi _{i}d\phi _{j}$ will be evaluated. We first
calculate the function in the diagonal part:
\begin{eqnarray}
&&h^{-2}S_{i}  \notag \\
&=&\frac{H_{i}}{\sqrt{\Delta }}H_{i}^{-1}r^{2}\Delta s_{i}+O(y^{2})  \notag
\\
&=&\frac{H_{i}}{\sqrt{\Delta }}\left( 1-\frac{q(C_{i}-1)}{r^{2}}\right) r^{2}%
\frac{a^{2}R^{4}}{r^{4}q}\left( 1-\frac{q(\alpha _{2}+2)}{aR^{2}}\right)
\frac{qr_{i}^{2}}{aR^{4}}\left( 1+\frac{q(\alpha _{2}+1)}{aR^{2}}\right)
+O(1/r^{4})+O(y^{2})  \notag \\
&=&\frac{H_{i}}{\sqrt{\Delta }}\frac{ar_{i}^{2}}{r^{2}}\left( 1-\frac{%
q(C_{i}-1)}{r^{2}}\right) \left( 1-\frac{q(\alpha _{2}+2)}{aR^{2}}\right)
\left( 1+\frac{q(\alpha _{2}+1)}{aR^{2}}\right) +O(1/r^{4})+O(y^{2})  \notag
\\
&=&\frac{H_{i}}{\sqrt{\Delta }}\mu _{i}^{2}\left( 1+\frac{qC_{i}}{r^{2}}%
\right) \left( 1-\frac{q(C_{i}-1)}{r^{2}}\right) \left( 1-\frac{q(\alpha
_{2}+2)}{r^{2}}\right) \left( 1+\frac{q(\alpha _{2}+1)}{r^{2}}\right)
+O(1/r^{4})+O(y^{2})  \notag \\
&=&\frac{H_{i}}{\sqrt{\Delta }}\mu _{i}^{2}+O(1/r^{4})+O(y^{2}).
\end{eqnarray}%
We next calculate the mixing part:
\begin{eqnarray}
&&h^{-2}h_{12}  \notag \\
&=&\frac{1}{\sqrt{\Delta }}\sqrt{\Delta }h^{-2}N_{12}  \notag \\
&=&\frac{1}{\sqrt{\Delta }}\frac{1}{4r^{2}K_{2}}N_{12}+O(y^{2})  \notag \\
&=&\frac{1}{\sqrt{\Delta }}\frac{1}{r^{2}}\frac{a^{2}R^{4}}{q}\frac{%
r_{1}^{2}r_{2}^{2}}{a^{2}R^{8}}\frac{q^{3}}{aR^{2}}(\alpha _{2}+\alpha
_{0}-\kappa _{1}-\kappa _{2})+O(1/r^{6})+O(y^{2})  \notag \\
&=&\frac{1}{\sqrt{\Delta }}\frac{1}{r^{2}}\frac{q^{2}r_{2}^{2}r_{3}^{2}}{%
aR^{6}}(\alpha _{2}+\alpha _{0}-\kappa _{1}-\kappa _{2})+O(1/r^{6})+O(y^{2})
\notag \\
&=&\frac{1}{\sqrt{\Delta }}\frac{q^{2}\mu _{1}^{2}\mu _{2}^{2}}{r^{4}}%
(\alpha _{2}+\alpha _{0}-\kappa _{1}-\kappa _{2})+O(1/r^{6})+O(y^{2}).
\end{eqnarray}

The metric in the angles is then expressed by \vspace{1pt}
\begin{eqnarray}
&&\sqrt{\Delta }h^{-2}h_{ab}(d\phi _{a}+M_{\phi _{a}}dt)(d\phi _{b}+M_{\phi
_{b}}dt)  \notag \\
&=&H_{1}\left[ \mu _{1}^{2}\left( 1+O(1/r^{4})\right) (d\phi _{1}+M_{\phi
_{1}}dt)^{2}\right]  \notag \\
&&+H_{2}\left[ \mu _{2}^{2}\left( 1+O(1/r^{4})\right) (d\phi _{2}+M_{\phi
_{2}}dt)^{2}\right]  \notag \\
&&+\left( \frac{2q^{2}\mu _{1}^{2}\mu _{2}^{2}}{r^{4}}(\alpha _{2}+\alpha
_{0}-\kappa _{1}-\kappa _{2})+O(1/r^{6})\right) (d\phi _{1}+M_{\phi
_{1}}dt)(d\phi _{2}+M_{\phi _{2}}dt).  \notag \\
&&
\end{eqnarray}


\quad


\vspace{1pt}

\subsection{Related formulas}

\vspace{1pt}

\vspace{1pt}\renewcommand{\theequation}{E.\arabic{equation}} %
\setcounter{equation}{0}

\vspace{1pt}

\label{formulaads}

\vspace{1pt}

\vspace{1pt}

\vspace{1pt}

In this appendix, we will summarize some formulas. These formulas
are also related to some useful expressions in \cite{Chen:2007du}.
(The conventions here are obtained from the conventions in
\cite{Chen:2007du} by $\Delta
\rightarrow (H_{2}H_{3})^{-2/3}\Delta $, the subscript change ($%
1,2,3\rightarrow 3,1,2$), $\rho _{i}\rightarrow R_{i}$.) We also summarize
equations which correspond to the two-charge superstar.
\begin{eqnarray}
&& h^{-2}=\frac{r^{2}\Delta+\mu_{3}^{2}}{\sqrt{\Delta}}, \\
&& \omega_{\phi_{i}}=-h^{2}\frac{\mu_{i}^{2}}{\sqrt{\Delta}}
,\quad (i=1,2)
\end{eqnarray}
\begin{equation}
S_{ij} = 2e^{i(\phi_{i}-\phi_{j})} \frac{\partial_{i}\bar{\partial}_{j}K}{Z+%
\frac{1}{2}} =\left\{
\begin{array}{l}
\frac{\mu_{i}^{2}+\sqrt{\Delta}h^{-2}H_{i}}{R_{i}^2\Delta} , \quad (i=j) \\
\frac{\mu_i \mu_j}{R_{i}R_{j}\Delta} , \quad (i\neq j)%
\end{array}%
\right.
\end{equation}%
where $H_{i}=1+q_i/r^2$. $R_{1}$ and $R_{2}$ obey a set of differential
equations:
\begin{eqnarray}
T_{1}=rR_{1}H_{2}f^{-1},  \notag \\
T_{2}=rR_{2}H_{1}f^{-1},  \label{Rdiffeq}
\end{eqnarray}
where $f=1+r^2H_{1}H_{2}$ and $T_{i}=dR_{i}/dr$.

For superstar, the $S_{i}$, $N_{12}$ and $M_{i}$ ($i=1,2$) are evaluated as
\begin{equation}
S_{i}=\frac{h^{2}\mu _{i}^{2}}{\sqrt{\Delta }}H_{i},~~~~~N_{12}=0,
\end{equation}%
\begin{equation*}
M_{i}=-1+H_{i}^{-1}=-\frac{q_{i}}{r^{2}+q_{i}}.
\end{equation*}

Formulas for AdS are available by taking $\Delta =1$ and $q_{i}=0$. The
differential equations (\ref{Rdiffeq}) can be solved to give
\begin{equation}
R_{1}=R_{2}=\sqrt{r^{2}+1}.
\end{equation}


\vspace{1pt}

\vspace{1pt}

\vspace{1pt}

\vspace{1pt}


\begin{thebibliography}{99}
\bibitem{Maldacena:1997re} J.~M.~Maldacena,
Adv.\ Theor.\ Math.\ Phys.\ \textbf{2}, 231-252 (1998)
[hep-th/9711200].


\bibitem{Gubser:1998bc} S.~S.~Gubser, I.~R.~Klebanov, A.~M.~Polyakov,
Phys.\ Lett.\ \textbf{B428}, 105-114 (1998) [hep-th/9802109].


\bibitem{Witten:1998qj} E.~Witten,
Adv.\ Theor.\ Math.\ Phys.\ \textbf{2}, 253-291 (1998)
[hep-th/9802150].


\bibitem{Corley:2001zk} S.~Corley, A.~Jevicki and S.~Ramgoolam,
Adv.\ Theor.\ Math.\ Phys.\ \textbf{5}, 809 (2002)
[arXiv:hep-th/0111222].


\bibitem{Berenstein:2004kk} D.~Berenstein,
JHEP \textbf{0407}, 018 (2004) [arXiv:hep-th/0403110].


\bibitem{Lin:2004nb} H.~Lin, O.~Lunin and J.~M.~Maldacena,
JHEP \textbf{0410}, 025 (2004) [arXiv:hep-th/0409174].


\bibitem{Berenstein:2005aa} D.~Berenstein,
JHEP \textbf{0601}, 125 (2006) [arXiv:hep-th/0507203 [hep-th]].

\bibitem{Balasubramanian:2005mg} V.~Balasubramanian, J.~de Boer, V.~Jejjala
and J.~Simon,
JHEP \textbf{0512}, 006 (2005) {[}arXiv:hep-th/0508023{]}.

\bibitem{Donos:2006iy} A.~Donos,
Phys.\ Rev.\ \textbf{D75}, 025010 (2007) [hep-th/0606199].


\bibitem{Chen:2007du} B.~Chen, \textit{et al.},
JHEP \textbf{0710}, 003 (2007) [arXiv:0704.2233 [hep-th]].


\bibitem{Lunin:2008tf} O.~Lunin, 
JHEP \textbf{0809}, 028 (2008) [arXiv:0802.0735 [hep-th]].

\bibitem{Chong:2004ce} E.~Gava, G.~Milanesi, K.~S.~Narain, M.~O'Loughlin,
JHEP \textbf{0705}, 030 (2007) [hep-th/0611065]; J.~P.~Gauntlett,
N.~Kim,
D.~Waldram, 
JHEP \textbf{0704}, 005 (2007) [hep-th/0612253]; Z.~-W.~Chong,
H.~Lu,
C.~N.~Pope, 
Phys.\ Lett.\ \textbf{B614}, 96-103 (2005) [hep-th/0412221];
N.~Kim,
JHEP \textbf{0601}, 094 (2006) [hep-th/0511029].


\bibitem{Lin:2010nd} H.~Lin,
arXiv:1008.5307 [hep-th].


\bibitem{Kimura:2007wy} Y.~Kimura, S.~Ramgoolam,
JHEP \textbf{0711}, 078 (2007) [arXiv:0709.2158 [hep-th]].

\bibitem{Brown:2007xh} T.~W.~Brown, P.~J.~Heslop, S.~Ramgoolam,
JHEP \textbf{0802}, 030 (2008) [arXiv:0711.0176 [hep-th]].

\bibitem{Bhattacharyya:2008rb} R.~Bhattacharyya, S.~Collins, R.~d.~M.~Koch,
JHEP \textbf{0803}, 044 (2008) [arXiv:0801.2061 [hep-th]].

\bibitem{flexp} H. Lin and J. P. Shock, unpublished.

\bibitem{Grant:2005qc} L.~Grant, L.~Maoz, J.~Marsano, K.~Papadodimas,
V.~S.~Rychkov,
JHEP \textbf{0508}, 025 (2005) [hep-th/0505079]; G.~Mandal,
JHEP \textbf{0508}, 052 (2005) [hep-th/0502104]; Y.~Takayama and
A.~Tsuchiya,
JHEP \textbf{0510}, 004 (2005) [arXiv:hep-th/0507070];
M.~M.~Caldarelli, D.~Klemm, P.~J.~Silva,
Class.\ Quant.\ Grav.\ \textbf{22}, 3461-3466 (2005)
[hep-th/0411203].


\bibitem{Horava:2005pv} P.~Horava, P.~G.~Shepard,
JHEP \textbf{0502}, 063 (2005) [hep-th/0502127].

\bibitem{Bhattacharyya:2008xy} R.~Bhattacharyya, R.~de Mello Koch and
M.~Stephanou, 
JHEP \textbf{0806}, 101 (2008) [arXiv:0805.3025 [hep-th]].

\bibitem{Brown:2008ij} T.~W.~Brown, P.~J.~Heslop and S.~Ramgoolam,
JHEP \textbf{0904}, 089 (2009) [arXiv:0806.1911 [hep-th]].

\bibitem{Kimura:2008ac} Y.~Kimura, S.~Ramgoolam,
Phys.\ Rev.\ D \textbf{78}, 126003 (2008) [arXiv:0807.3696
[hep-th]].

\bibitem{Kimura:2010tx} Y.~Kimura,
JHEP \textbf{1005}, 103 (2010) [arXiv:1002.2424 [hep-th]].

\bibitem{Koch:2008cm} R.~d.~M.~Koch, N.~Ives and M.~Stephanou,
Phys.\ Rev.\ D \textbf{79}, 026004 (2009) [arXiv:0810.4041
[hep-th]].

\bibitem{Behrndt:1998ns} K.~Behrndt, A.~H.~Chamseddine, W.~A.~Sabra,
Phys.\ Lett.\ \textbf{B442}, 97-101 (1998) [hep-th/9807187];
K.~Behrndt, M.~Cvetic, W.~A.~Sabra,
Nucl.\ Phys.\ \textbf{B553}, 317-332 (1999) [hep-th/9810227].

\bibitem{Myers:2001aq} R.~C.~Myers, O.~Tafjord,
JHEP \textbf{0111}, 009 (2001) [hep-th/0109127].

\bibitem{Chen:2007gh} H.~-Y.~Chen, D.~H.~Correa, G.~A.~Silva,
Phys.\ Rev.\ \textbf{D76}, 026003 (2007) [hep-th/0703068
[hep-th]].


\bibitem{Koch:2008ah} R.~d.~M.~Koch, 
JHEP \textbf{0811}, 061 (2008) [arXiv:0806.0685 [hep-th]].


\bibitem{Lin:2010sba} H.~Lin, A.~Morisse, J.~P.~Shock,
JHEP \textbf{1006}, 055 (2010) [arXiv:1003.4190 [hep-th]].


\bibitem{Kimura:2009jf} Y.~Kimura,
JHEP \textbf{0912}, 044 (2009) [arXiv:0910.2170 [hep-th]].


\bibitem{Kimura:2009ur} Y.~Kimura, S.~Ramgoolam and D.~Turton,
JHEP \textbf{1005}, 052 (2010) [arXiv:0911.4408 [hep-th]].

\bibitem{Brown:2010pb} T.~W.~Brown,
JHEP \textbf{1005}, 058 (2010) [arXiv:1002.2099 [hep-th]].

\bibitem{Koch:2011hb} R.~d.~M.~Koch, M.~Dessein, D.~Giataganas and
C.~Mathwin, 
arXiv:1108.2761 [hep-th]. 


\bibitem{Pasukonis:2010rv} J.~Pasukonis and S.~Ramgoolam,
JHEP \textbf{1102}, 078 (2011) [arXiv:1010.1683 [hep-th]].

\bibitem{DeComarmond:2010ie} V.~De Comarmond, R.~de Mello Koch and
K.~Jefferies, 
JHEP \textbf{1102}, 006 (2011) [arXiv:1012.3884 [hep-th]].


\bibitem{Carlson:2011hy} W.~Carlson, R.~d.~M.~Koch, H.~Lin,
JHEP \textbf{1103}, 105 (2011) [arXiv:1101.5404 [hep-th]].
\end{thebibliography}
\end{document}